\documentclass[a4paper,11pt]{article}
\pdfoutput=1 %

\usepackage{amsmath}
\usepackage{amssymb}

\newcommand{\fms}[1]{{#1}\!\!\!\slash}

\newcommand{\nb}{\overline{n}}

\newcommand{\nbs}{\fms{\overline{n}}}

\newcommand{\ie}{i\epsilon}

\newcommand{\braB}{\,\langle B \!\mid\,}
\newcommand{\ketB}{\,\mid\! B \rangle\,}

\newcommand{\nn}{\nonumber}

\usepackage{jheppub} 

\usepackage{slashed}

\title{\boldmath Subleading Power Factorization in $\bar B\rightarrow X_s \ell^+ \ell^-$}

\author[a]{Michael Benzke,}
\author[b]{Tobias Hurth}
\author[b]{and Sascha Turczyk}

\affiliation[a]{II. Institute for Theoretical Physics, University Hamburg\\ 
Luruper Chaussee 149, D-26761 Hamburg, Germany}
\affiliation[b]{PRISMA Cluster of Excellence \& Mainz Institut for Theoretical
Physics, Johannes Gutenberg University, 55099 Mainz, Germany}

\emailAdd{{michael.benzke@desy.de,hurth@uni-mainz.de,turczyk@uni-mainz.de}}

\abstract{We analyze the factorization to subleading power in the flavor changing neutral current process $\bar B\rightarrow X_s \ell^+ \ell^-$. In particular, we compute the so-called resolved contributions and explore the numerical impact on observables. In these contributions the virtual photon couples to light partons instead of connecting directly to the effective weak-interaction vertex. They represent an irreducible uncertainty in the  inclusive  $\bar B\rightarrow X_s \ell^+ \ell^-$ decay  which cannot  be removed by relaxing the experimentally necessary cuts in the hadronic mass spectrum.}

\begin{document} 
\begin{flushright}
MITP/17-019 \\
\end{flushright}
\maketitle

\newpage
\section{Introduction}

Flavour changing neutral currents (FCNC) play an essential role in the search for New Physics (NP) effects. The leading order Standard Model (SM) process already occurs only at the loop-level and consequently any new physics (NP) effects beyond the SM may enter at the same level.  However, up to some $2-3 \sigma$ deviations in FCNC 
no signal of NP has been detected yet. 
Due to this current lack of really significant deviations from SM predictions, any NP is either out of reach of the current colliders or has a peculiar flavour structure.   
This is the famous flavour problem, i.e. the question why FCNCs  are suppressed (for a review see Ref.~\cite{Isidori:2010kg}).
 This problem must be solved in any viable NP model. In both options, a thorough investigation of the flavour structure is mandatory in order to explore the underlying NP model.

The inclusive decay mode $\bar  B \to X_s \ell^+\ell^-$ is one of the most important, theoretically clean modes of the indirect search for new physics via flavour observables (for reviews see Refs.~\cite{Hurth:2003vb,Hurth:2010tk,Hurth:2012vp}). 
Compared   with the $\bar B \rightarrow X_s \gamma$ decay,
the inclusive $\bar B \rightarrow X_s \ell^+ \ell^-$ decay presents a
complementary and more complex test of the SM, given that  different
perturbative electroweak contributions add to the decay rate. As a three body decay process it also offers more observables. Due to the presence of the lepton-antilepton pair, more structures contribute to the decay rate and some subtleties in the formal theoretical description arise which one  needs to scrutinize. 
It is generally assumed that  this  inclusive mode is dominated by perturbative contributions like the inclusive $\bar B \to X_s \gamma$ decay if  one eliminates $c \bar c$ resonances with the help of kinematic cuts.
In the so-called perturbative $q^2$ windows below and above the
resonances, namely in the low-dilepton mass region $1\;{\rm GeV}^2 < q^2
= m_{\ell\ell}^2 < 6\;{\rm GeV}^2$ as well as  in the high-dilepton mass
region where  $q^2 > 14.4\;{\rm GeV}^2$
these perturbative contributions are well explored and have already reached a highly sophisticated level. The most recent analysis of all angular observables in the $\bar B \rightarrow X_s \ell^+\ell^-$ decay was given 
in Ref.~\cite{Huber:2015sra}; it includes all available perturbative NNLO QCD, NLO QED corrections and also the {\it known}  subleading power corrections.

{The inclusive mode $\bar B\rightarrow X_s \ell^+ \ell^-$  allows for an important crosscheck of the recent LHCb  data  on the corresponding exclusive mode. The so-called anomalies found in some angular observables of the exclusive decay $B \to K^* \mu^+\mu^-$~\cite{Aaij:2013qta,Aaij:2015oid}
cannot be interpreted unambiguously because of  the unknown subleading power corrections in the theoretical framework of QCD improved factorization. One cannot decide at the moment if these deviations from the SM are 
first signs for new physics beyond the SM,  a consequence of the unknown hadronic power corrections or just statistical fluctuations. 
As was shown in Refs.~\cite{Hurth:2013ssa,Hurth:2014zja}, the future measurements of the inclusive mode will be able to resolve this puzzle.}

{Belle and BaBar have measured the branching ratio using the sum-of-exclusive technique only. Unfortunately, the latest published measurement of Belle~\cite{Iwasaki:2005sy} is based on less than $30\%$ of the data set 
available at the end of the Belle experiment, i.e. on  a sample of $152 \times 10^6$ $B \bar B$ events only.
At least  BaBar has published an analysis based on the whole data set of Babar using a sample of $471 \times 10^6$ $B \bar B$ events~\cite{Lees:2013nxa} which updated the former analysis of 2004~\cite{Aubert:2004it}. However,  Belle has already measured the forward-backward asymmetry~\cite{Sato:2014pjr}, while  BaBar presented a measurement of the CP violation in this channel~\cite{Lees:2013nxa}.  All these  measurements are  still limited by the statistical errors.  
The super flavour factory Belle~II at KEK will accumulate data samples that are two orders of magnitude larger~\cite{Belle2}. This will push experimental precision to its limit. Thus, also a precise understanding of the theoretical predictions is necessary. }

{The inclusive modes $B \rightarrow X_s \gamma$ and $B
\rightarrow X_s \ell^+ \ell^-$ are dominated by the partonic contributions which can be calculated perturbatively. 
It is well-known that  the heavy mass expansion (HME)  makes it possible  to calculate
the inclusive decay rates of a hadron containing a heavy quark, {\it if} only the leading operator in the effective
Hamiltonian (${\cal O}_7$ for $B \to X_s \gamma$, ${\cal O}_9$ for $B \to X_s \ell^+\ell^-$) 
is considered~\cite{Chay:1990da,Bigi:1992su}.  The HME represents a local operator product expansion (OPE) based on the optical theorem. The partonic contribution is the leading term in this expansion in power of $1/m_b$.   
Due to the equations of motion, there is no contribution of order $\Lambda/m_b$. Thus, the corrections to the partonic contribution start with $1/m_b^2$ only and have a  rather small numerical impact. For the inclusive decay  $\bar B \to X_s \ell^+ \ell^-$ 
these leading hadronic power corrections  with $1/m_b^2$ and $1/m_b^3$ have already been analysed in  Refs.~\cite{Falk:1993dh, Ali:1996bm, Chen:1997dj, Buchalla:1998mt, Bauer:1999kf} (for the inclusive decay $\bar B \to X_s \gamma$ see Ref.~\cite{Mannel:2010wj}).}

{However, as already noted in Ref.~\cite{Ligeti:1997tc}, there is no OPE for the
inclusive decay $B \rightarrow X_s \gamma$ if one includes operators beyond the leading electromagnetic dipole operator ${\cal
  O}_7$ into the analysis.  Voloshin~\cite{Voloshin:1996gw} has identified such
a contribution to the total decay rate in the interference of the $b \to s
\gamma$ amplitude due to the electromagnetic dipole operator ${\cal
  O}_7$ and the charming penguin amplitude due to the current-current
operator ${\cal O}_2$. It is described by matrix element of a non-local operator. This is an example of a so-called resolved photon contribution. Such a contribution  contains subprocesses in which the photon couples to light partons
instead of connecting directly to the effective weak-interaction vertex~\cite{Benzke:2010js}.\footnote{It is possible to expand this non-local  contribution to  local operators again if one assumes that the charm is a heavy quark. Then  the
first term in this expansion is the dominating one~\cite{Ligeti:1997tc,Grant:1997ec,Buchalla:1997ky}. This
non-perturbative correction is suppressed by $\lambda_2/m_c^2$ and is
estimated to be of order $3\%$ compared with the leading-order
(perturbative) contribution to $\Gamma_{b \to s \gamma}$.
But if one assumes that the charm mass scales as 
$m_c^2\sim\Lambda_{\text{QCD}}  m_b$, the charm penguin contribution must be
described by the matrix element of a non-local
operator~\cite{Benzke:2010js}.}}

{An analysis of all resolved photon contributions to the inclusive decay $\bar B \to X_s\gamma$
related to other operators in the weak
Hamiltonian has been presented in Ref.~\cite{Benzke:2010js} (see also Ref.~\cite{Lee:2006wn}).                         
All these non-local contributions manifestly demonstrate 
the breakdown of the local OPE within the
hadronic power corrections.  However, such non-local  power corrections lead to a multi-scale problem which can be analysed well within soft-collinear effective  theory (SCET).
These non-local matrix elements are very difficult to estimate. It has been shown that there is 
an irreducible theoretical uncertainty of $\pm
(4-5)\%$ for the total $CP$ averaged decay rate, defined with a
photon-energy cut of $E_\gamma = 1.6$ GeV~\cite{Benzke:2010js}.}

In the present  paper we explore the subleading power factorization  of  the inclusive decay $\bar B\rightarrow X_s \ell^+ \ell^-$ and its implications to observables. 
Within the inclusive decay {$\bar B \to X_s \ell^+ \ell^-$}, 
the hadronic ($M_X$) and dilepton invariant ($q^2$) masses are independent
kinematical quantities.  In order  to suppress potential huge backgrounds  one needs an invariant mass cut on the hadronic final state system ($M_X \lesssim 2\,\text{GeV}$). This cut poses no additional constraints in the high-dilepton mass region, but in the low-dilepton  one the cut on the hadronic mass implies specific  kinematics in which the standard OPE  breaks down and  non-perturbative b-quark distributions, so-called shape functions,  have to be introduced. The specific kinematics of low dilepton masses $q^2$ and of small hadronic masses $M_X$  leads to a multi-scale problem for which soft-collinear effective theory (SCET) is the appropriate tool.

{A former SCET analysis uses the universality of the leading shape
function to show that the reduction due to the $M_X$-cut    in all angular observables of  the inclusive decay $\bar B\rightarrow X_s \ell^+ \ell$
can be accurately computed. The  effects of subleading shape functions
lead to an additional uncertainty of $5\%$~\cite{Lee:2005pk,Lee:2005pw}.\footnote{In a later analysis~\cite{Lee:2008xc}
the uncertainties due to subleading shape functions are 
conservatively estimated.  Using the combined $B \to
X_s\gamma$, $B \to X_u\ell \bar\nu$ and $B \to X_s \ell^+\ell^-$
data the uncertainties due to leading and sub-leading shape functions can be reduced in the future~\cite{Lee:2008xc}.}
However, in all these previous analyses a problematic assumption is made, namely  that  $q^2$ represents a hard scale in the kinematical region of low $q^2$ and of small $M_X$.
As we will show explicitly in our present  SCET analysis, the hadronic cut  implies  the scaling of $q^2$ being  not hard but (anti-) hard-collinear in the low-$q^2$ region.}

The main goal of the paper is to identify  the correct power counting of all the variables in the low-$q^2$ window of the inclusive decay $\bar B \rightarrow X_s \ell^+\ell^-$  within the effective theory SCET 
if a hadronic mass  cut is imposed. Furthermore we will  analyse the resolved power corrections in a systematic way  and present numerical estimates of the corresponding uncertainties.
As mentioned above, in these contributions the virtual photon couples to light partons instead of connecting directly to the effective weak-interaction vertex.  Moreover, we will show that the resolved  contributions - as a special feature  -  stay non-local when the hadronic mass cut is released. In this sense they represent an irreducible uncertainty independent of the hadronic mass cut.

The paper is organized as follows. 
In section~\ref{sec:theory} we introduce  the theoretical framework, in particular we identify the correct power counting and the factorization properties of the subleading contributions.  In section~\ref{sec:differential} we derive the fully differential decay rate. In section~\ref{sec:example}  we present the explicit calculation  of the interference term of the ${\cal O}_7$ and the ${\cal  O}_2$ operators. In  Section~\ref{sec:contributions} we present the analytical results of all resolved contributions in the first subleading power.  Their numerical impact is   investigated in section~\ref{sec:numerics}. Finally we summarize and discuss the obtained results in 
section \ref{sec:conclusion}.

\newpage

\section{Theoretical Framework}   \label{sec:theory}

The effective operator basis for the underlying parton interaction of the semi-leptonic flavour changing neutral current decay $\bar B \to X_s \ell^+\ell^-$ is well-known~\cite{Buchalla:1995vs}.  Many higher-order calculations have led to the availability of NNLO precision and NNLL resummation in the strong coupling $\alpha_s$. At the relevant scale $m_b$ of the $b$-quark, all heavier fields are integrated out, and the effective operator basis contains only active flavours. In our convention, corresponding to the one used in~\cite{Beneke:2001at}, the contributing operators are given by
\begin{subequations}\label{eq:op_basis}
\begin{alignat}{2}
   {\cal O}_1^q &= (\bar q b)_{V-A} (\bar s q)_{V-A} & \qquad {\cal O}^q_2 &= (\bar q_i b_j)_{V-A} (\bar s_j q_i)_{V-A} \,, \\
   {\cal O}_3 &= (\bar s b)_{V-A} \sum_{q}\,(\bar q q)_{V-A} & \qquad {\cal O}_4 &= (\bar s_i b_j)_{V-A} \sum_{q}\,(\bar q_j q_i)_{V-A} \,, \\
   {\cal O}_5 &= (\bar s b)_{V-A} \sum_{q}\,(\bar q q)_{V+A} & \qquad {\cal O}_6 &= (\bar s_i b_j)_{V-A} \sum_{q}\,(\bar q_j q_i)_{V+A} \,, \\
   {\cal O}_{7\gamma} &= -\frac{e}{8\pi^2}\,m_b\,
    \bar s\sigma_{\mu\nu}(1+\gamma_5) F^{\mu\nu} b & \qquad {\cal O}_{8g} &= -\frac{g_s}{8\pi^2}\,m_b\,
    \bar s\sigma_{\mu\nu}(1+\gamma_5) G^{\mu\nu} b\,, \\
   {\cal O}_9 &= \frac{\alpha}{2\pi} (\bar sb)_{V-A} (\bar \ell \ell)_{V}& \qquad  {\cal O}_{10} &= \frac{\alpha}{2\pi} (\bar sb)_{V-A} (\bar \ell\ell)_{A} \,,
\end{alignat}
\end{subequations}
with $q=u,c$ and $i,j$ denoting the color indices and $(\bar q_1 q_2)_{V\pm A} = \bar q_1 \gamma_\mu (1 \pm \gamma_5)  q_2$.   
Our sign convention is such that $iD_\mu=i\partial_\mu+g_s\,T^a A_\mu^a+e\,Q_f A_\mu$, where $T^a$ are the $SU(3)$ color generators, and $Q_f$ is the electric charge of the fermion in units of $e$. Using Standard Model CKM unitarity, with $\lambda_q=V_{qb} V_{qs}^*$ and $\lambda_u + \lambda_c + \lambda_t = 0$, we may write the effective Hamiltonian as
\begin{equation}\label{eq:WeakHamiltonian}
   {\cal H}_\text{eff} = \frac{G_F}{\sqrt{2}} \sum_{q=u,c} \lambda_q\,
   \bigg( C_1\,{\cal O}_1^q + C_2\,{\cal O}_2^q+ C_{7\gamma}\,{\cal O}_{7\gamma} + C_{8g}\,{\cal O}_{8g}  + \sum_{i=3,...,6,9,10} C_i\,{\cal O}_i 
   \bigg) \,.
\end{equation}
The Wilson coefficients $C_i$ depend on the scale $\mu$ at which the operators are renormalized and in our convention $C_{7\gamma}$ is negative.  Here the four-quark and QCD-penguin operators ${\cal O}_{1-6}$, and the QED and QCD dipole operators ${\cal O}_{7\gamma,8 g}$ can contribute via an appropriate contraction with the QED Lagrangian to the process in question.

\subsection{Set-up of the SCET ansatz}

Calculating the inclusive decay mode $\bar B \to X_s \ell^+\ell^-$ we face two problems. On the one hand, the integrated branching fraction is dominated by
resonant $q\bar q$ background, especially with $q=c$, i.e. resonant $J/\psi
\rightarrow \ell^+ \ell^-$ intermediate states for the (virtual) photon, which exceeds the non-resonant
charm-loop contribution by two orders of magnitude. This feature should not be misinterpreted as a striking failure of global parton-hadron duality as shown in Ref.~\cite{Beneke:2009az}.
However,   $c \bar c$ resonances that show up as large peaks in the dilepton invariant mass spectrum are removed by appropriate  kinematic cuts -- leading to so-called `perturbative $q^2$-windows', namely  the low-dilepton-mass  region $1\,{\rm GeV}^2 < q^2 = m_{\ell\ell}^2   < 6\,{\rm GeV}^2$, and also the high-dilepton-mass region with $q^2 > 14.4\,{\rm GeV}^2$.  

On the other hand, in a realistic experimental environment we need to suppress potential huge backgrounds by an invariant mass cut on the hadronic final state system ($M_X \lesssim 2\,\text{GeV}$). This cut poses no additional constraints in the high-dilepton-mass region. But in the
low-dilepton mass region we have in the $B$ meson rest frame due to $q= P_B -P_X$
\begin{equation} 
2\, M_B\, E_X \, = M_B^2 +M_X^2 -q^2\,.
\end{equation}
Thus, for low enough $q^2$ in combination with $M_X^2 \ll E_X^2$ the $X_s$ system is jet-like with  $E_X  \sim  M_B$. This further implies hat $P_X$ is near the light cone.

Within these kinematic constraints, soft-collinear-effective theory (SCET)~\cite{Bauer:2001yt} is the appropriate tool to study the factorization properties of inclusive $B$-meson decays and to 
analyse the multi-scale problem. The cuts in the two independent kinematic variables, namely the hadronic and dilepton invariant masses, force us  to study the process in the so-called shape function region
with a large energy $E_X$ of order   $M_B$ and low invariant mass  $ M_X \sim \sqrt{m_b \Lambda_\text{QCD}}$ of the hadronic system. SCET enables us to systematically obtain a scaling law of the momentum components. In our set-up the scales $\Lambda_\text{QCD}$, $M_X$, and $M_B$ are relevant. For the ratio of these scales, one finds the following hierarchy:
\begin{equation}
  \frac{\Lambda_\text{QCD}}{M_B} \ll \frac{M_X}{M_B} \ll 1\,.
\end{equation}
Hence, resumming logarithms between these scales becomes important. SCET allows to systematically resum the logarithms of these scale ratios, and more importantly factorizes the effects stemming from the different regions. This enables us to calculate the process in a consistent expansion, and to factorize off effects that can be calculated  perturbatively. This reduces the non-perturbative quantities to a limited set of soft functions. Defining $\lambda = \Lambda_\text{QCD}/M_B$, we numerically have $M_X \lesssim \sqrt{M_B \Lambda_\text{QCD}} \sim M_B \sqrt{\lambda}$. This sets the power-counting scale for the possible momentum components in light-cone coordinates $n^\mu = (1,0,0,1)$ and $\bar n^\mu = (1,0,0,-1)$.  Any four-vector may be decomposed  according to  $a^\mu = n\cdot a \,\,\bar n^\mu/2 + \bar n\cdot a \,\, n^\mu/2 + a_\perp^\mu\,.$
We use the short-hand notation $a \sim (n\cdot a, \bar n\cdot a, a_\perp)$ to indicate the scaling of the momentum components in powers of $\lambda$. Within the validity of SCET,  we have a hard momentum region $p_\text{hard} \sim (1, 1, 1)$, a hard-collinear region $p_\text{hc} \sim ( \lambda, 1 , \sqrt{\lambda})$, an anti-hard-collinear region $p_{\overline{\text{hc}}} \sim ( 1, \lambda , \sqrt{\lambda})$ and a soft region $p_\text{soft} \sim (\lambda, \lambda, \lambda)$. 

\begin{figure}[hpt]
    \centering\includegraphics[scale=0.36]{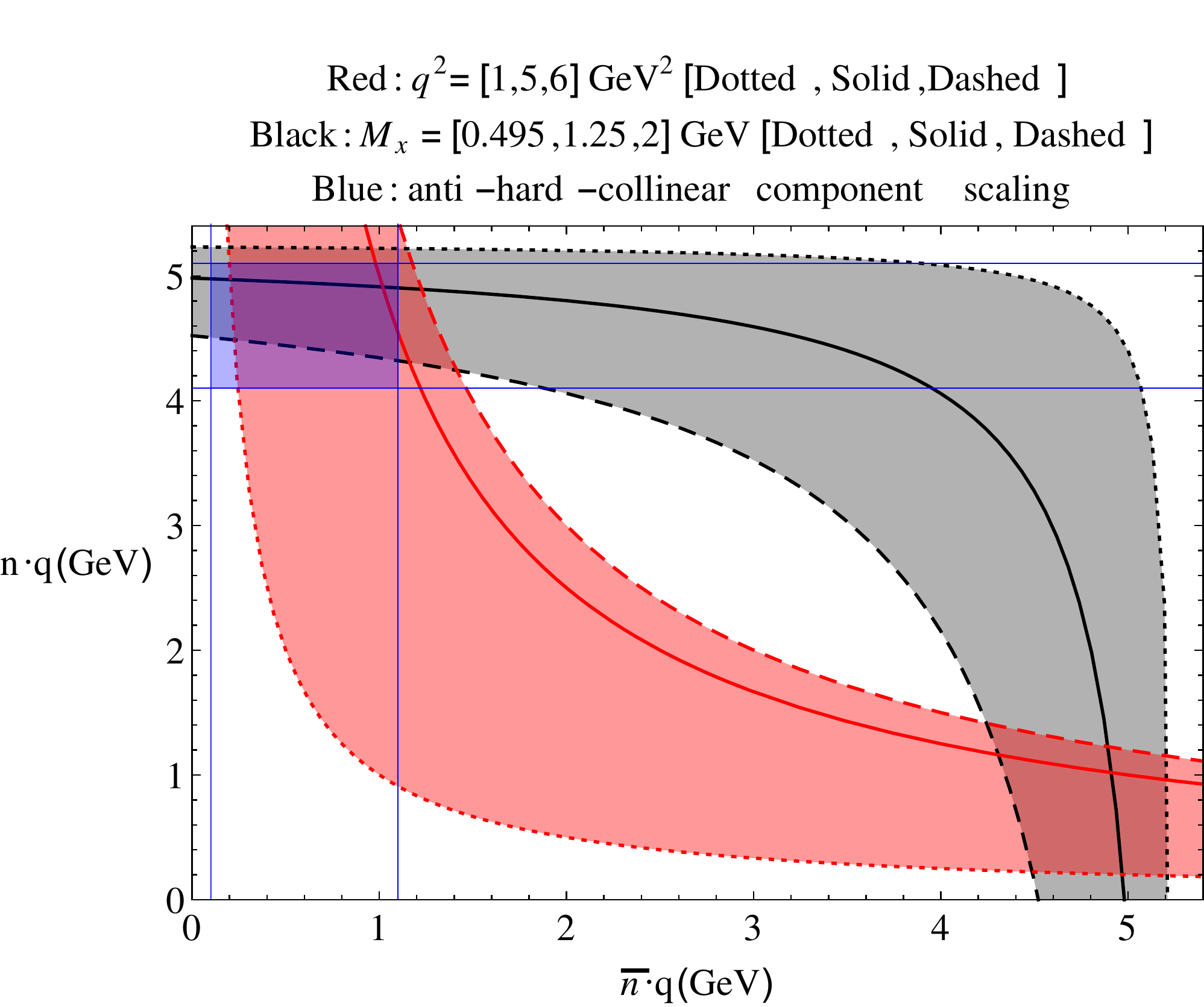}\hfill\includegraphics[scale=0.36]{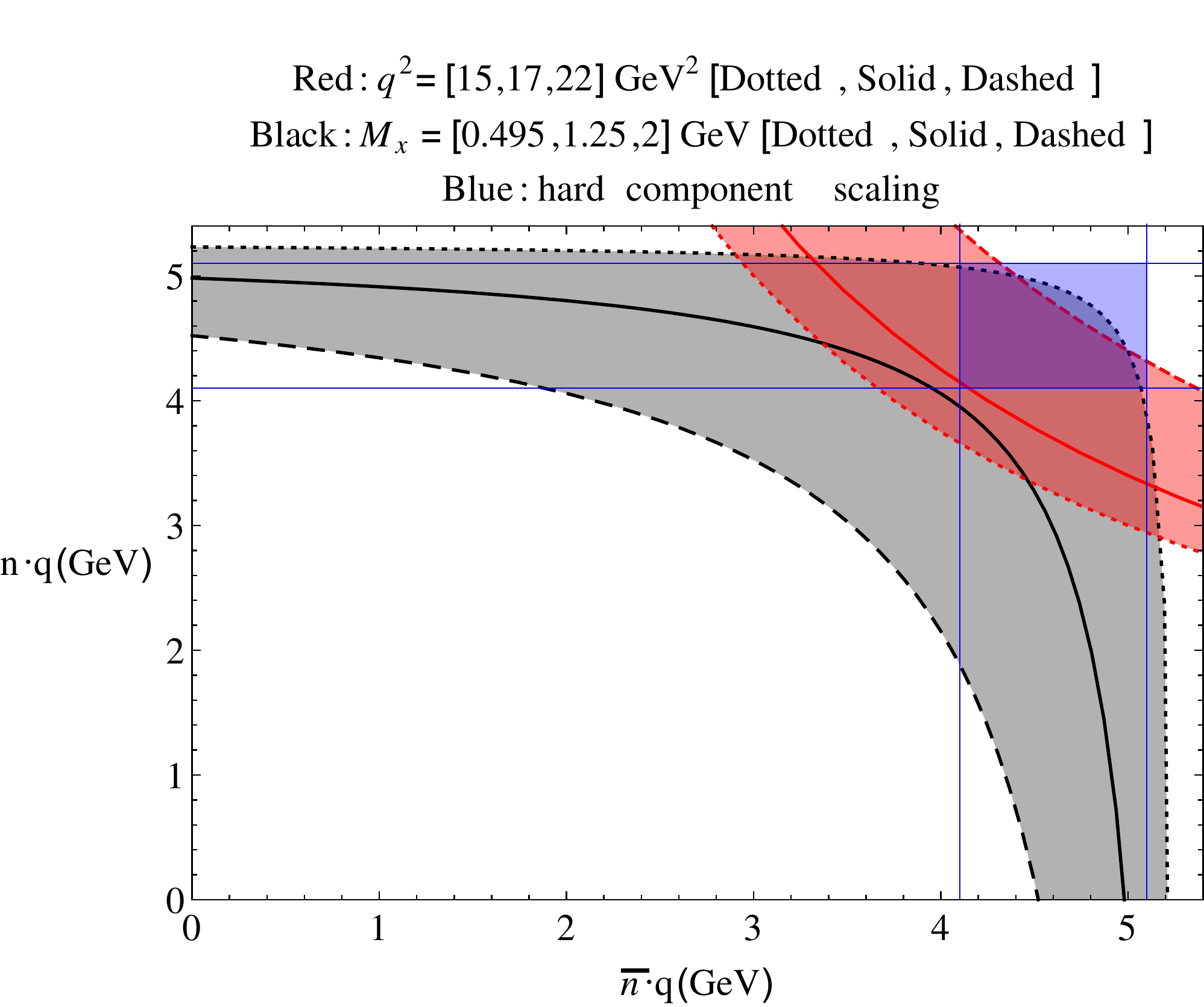}
    \caption{$q^2 = (n \cdot q) (\bar n \cdot q)$ with $q_\perp = 0$ for the two perturbative mass windows. The gray band shows the experimental hadronic invariant mass cut with the $K$ as the lowest mass state, and the red band corresponds to the $q^2$ cut. The blue lines indicate the scaling of the two light-cone components. Left: Low invariant mass window. Scaling of $q_{\overline{\text{hc}}}$ is indicated. Right: High invariant mass window, with the maximally allowed value of $M_B$. Scaling of $q_\text{hard}$ is indicated.}\label{Fig:scaling_law}
\end{figure}

As far as the two-body radiative decay is concerned, the kinematics  imply  $q^2 = 0$ and $E_\gamma \sim m_b/2$, and, taking into account the invariant mass and photon energy requirements, the scaling is fixed to be a hard-collinear hadronic jet recoiling against an anti-hard-collinear photon. 

In the case of a lepton-antilepton pair in the final state, we need to restrict the momentum transfer to the leptons outside the mass window of the $c\bar c$ resonances as described above. In Fig.~\ref{Fig:scaling_law} we compare the momentum scaling of the lepton-antilepton pair in terms of the light-cone coordinate decomposition and the experimental cuts. The gray band corresponds to the hadronic invariant mass cut in order to suppress background, while the red band is the $q^2$ constraint to reject the $c\bar c$ resonances. The blue lines show the validity of SCET in terms of the momentum component scaling, on the left figure for an anti-hard-collinear scaling, while on the right one for a hard momentum scaling. Note that there exist two  solutions for the left figure, as we may view the leptons to be anti-hard-collinear and the hadronic jet collinear and vice versa.
Obviously, the high mass window corresponds to hard leptons and is outside of the validity of a description in terms of SCET. It can be readily seen that the current mass cuts do not have an impact on this scenario. That is in contrast to the low $q^2$ region. The overlap of the red and gray band is the allowed region after experimental cuts, and it is in good agreement with our assumptions for the effective theory, which is approximately given by the blue rectangle. Therefore with assigning an anti-hard-collinear momentum to the virtual photon and a hard-collinear one to the hadronic system, we are in a good approximation in the validity window of both the experimental requirement and the effective theory. 

\begin{figure}[hpt]
    \centering\includegraphics[scale=0.36]{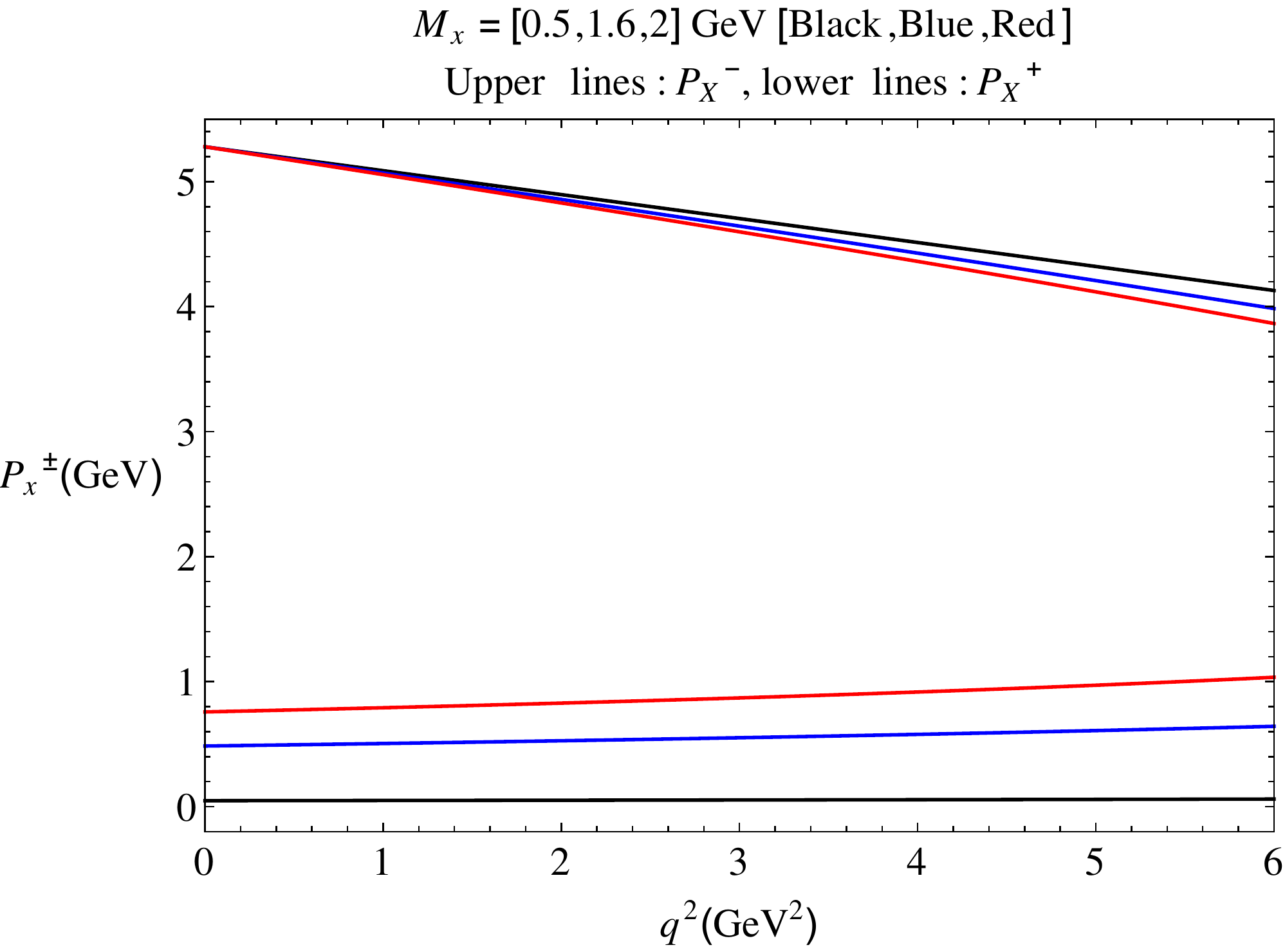}\hfill\includegraphics[scale=0.36]{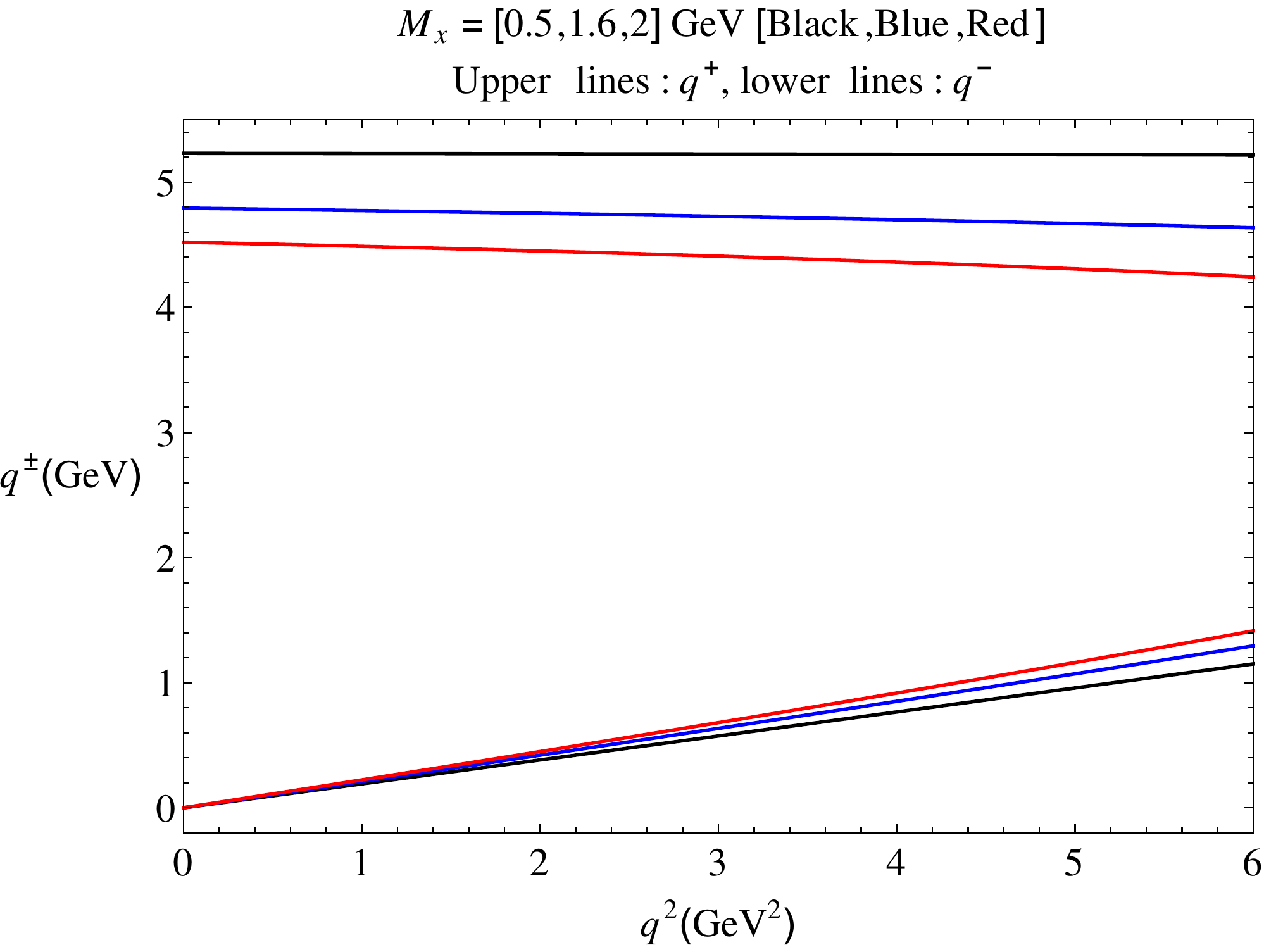}
    \caption{The scaling of the momentum components of the hadronic system $P_X^+ = n\cdot P_X$ and $P_X^- = \bar n \cdot P_X$ (left) and the lepton system $q^+ = n\cdot q$ and $q^- = \bar n \cdot q$ is plotted as a function of $q^2$ each for three values of the hadronic invariant mass.}\label{Fig:scaling_law_momentum}
\end{figure}

To show this more explicitly, we can introduce the two light-cone components of the hadronic momentum  with  $n\cdot P_X\,\bar n\cdot P_X = M^2_X$ and $P_X^\perp = 0$ 
\begin{equation}
\bar n\cdot P_X =   E_X  +  |  \vec{P}_X |  \sim   {O}(M_B)\,\,,\,  n\cdot P_X =   E_X  -  | \vec{P}_X |  \sim   {O}(\Lambda_{\rm QCD})\,.
\end{equation}
Using the kinematical relations, the leptonic light-cone variables are given by 
\begin{equation}
n\cdot q = M_B - n\cdot P_X\,,\, \bar n\cdot q = M_B - \bar n\cdot P_X = q^2 / (M_B - n\cdot P_X)\,.
 \end{equation}
In  Fig.~\ref{Fig:scaling_law_momentum}, we show the scaling  of the momentum components of the hadronic system $n\cdot P_X$ and $\bar n \cdot P_X$  (left plot) 
and of the lepton system  $n\cdot q$ and $\bar n \cdot q$ (right plot)   as function of $q^2$ for three different values of the hadronic  mass cut.  
It can be clearly seen, that for the experimentally invoked cuts without further assumptions other than assuming the effective two-body decay system $B\rightarrow X_s \gamma^*$ to be aligned along the light-cone axis without a perpendicular component, the hadronic system scales as hard-collinear, while the lepton system scales as anti-hard collinear.
However, as also can be extracted from these plots, a lower cut of $q^2 \lesssim 5 \text{ GeV}^2$ instead of $q^2 \lesssim 6 \text{ GeV}^2$ is preferred  because a higher value of the $q^2$ cut pushes the small component to values slightly beyond our assumptions of the momentum component scaling and therefore neglected higher order terms may have a more sizable contribution. Nevertheless, the  assumption of a hard $q$ momentum as used 
in the calculations of Refs.~\cite{Lee:2005pk,Lee:2005pw,Lee:2008xc} is not appropriate. Moreover, it implies a different scaling and also a different matching of the operators. And as we will show below, this assumption  would  imply that there are no resolved contributions in the effective field theory.

\subsection{Factorization theorem and operator matching and scaling}

\begin{figure}
\begin{center} 
\epsfig{file=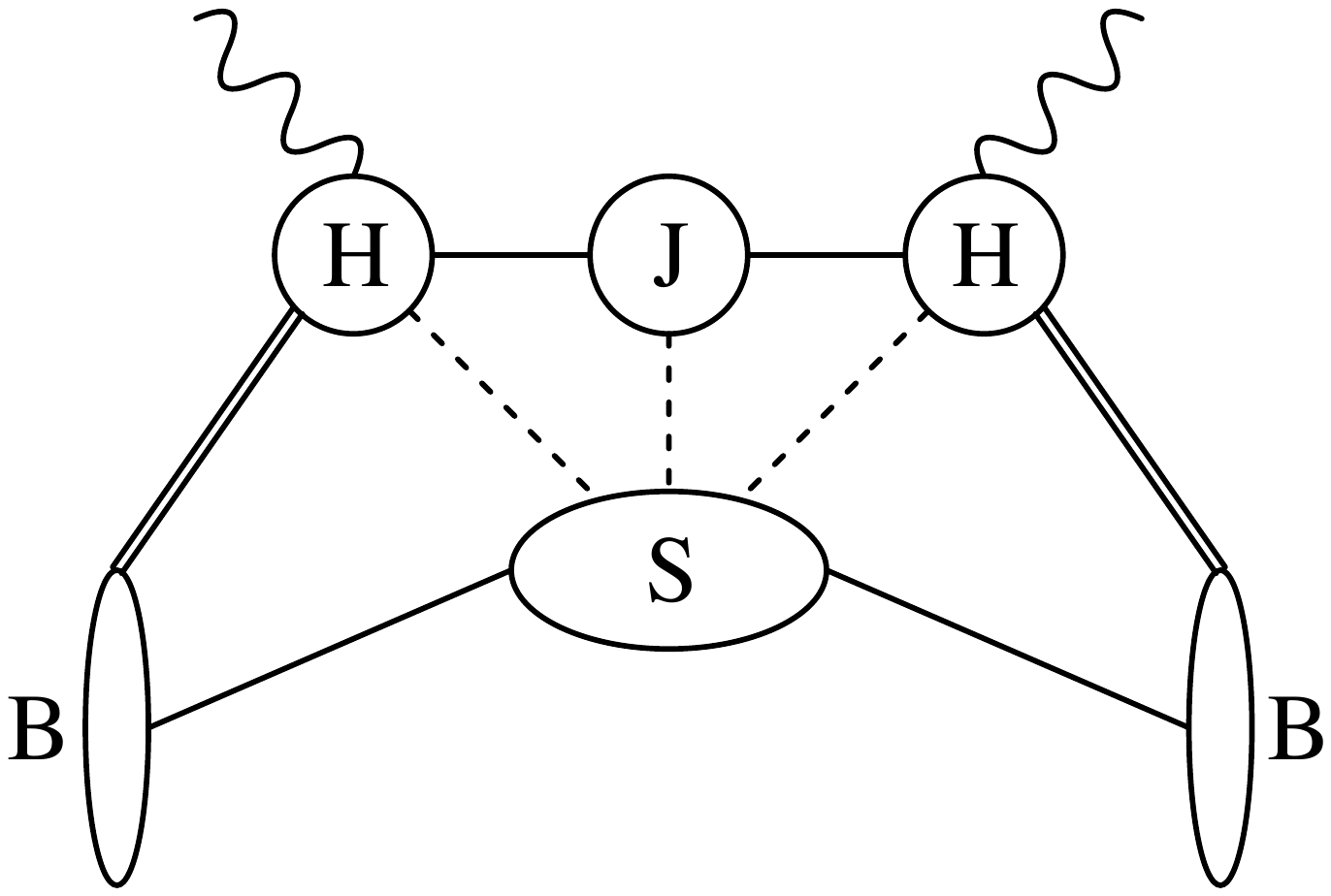,width=5.9cm}\hspace{-1.5cm}\epsfig{file=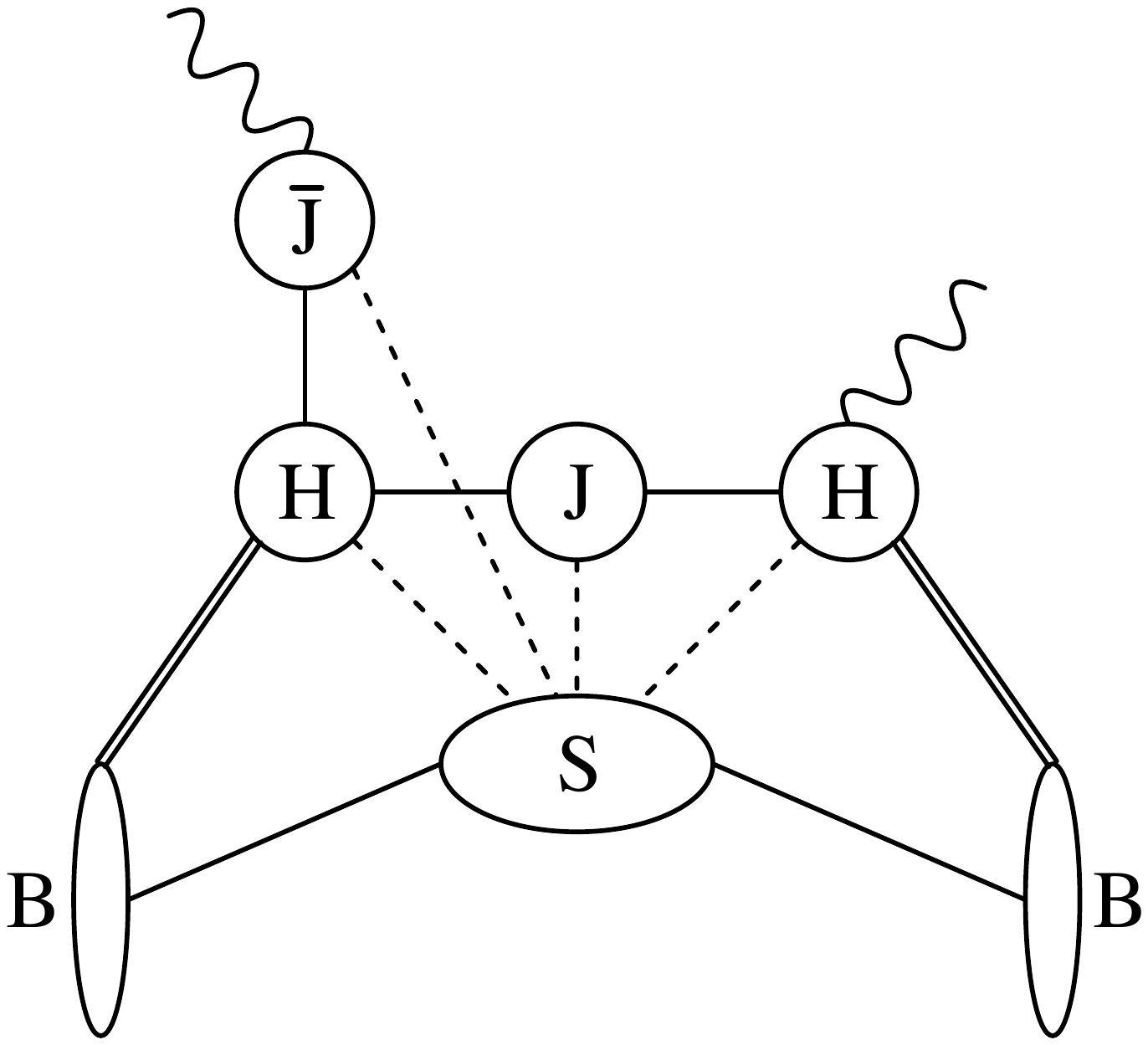,width=5.9cm}\hspace{-1.5cm}\epsfig{file=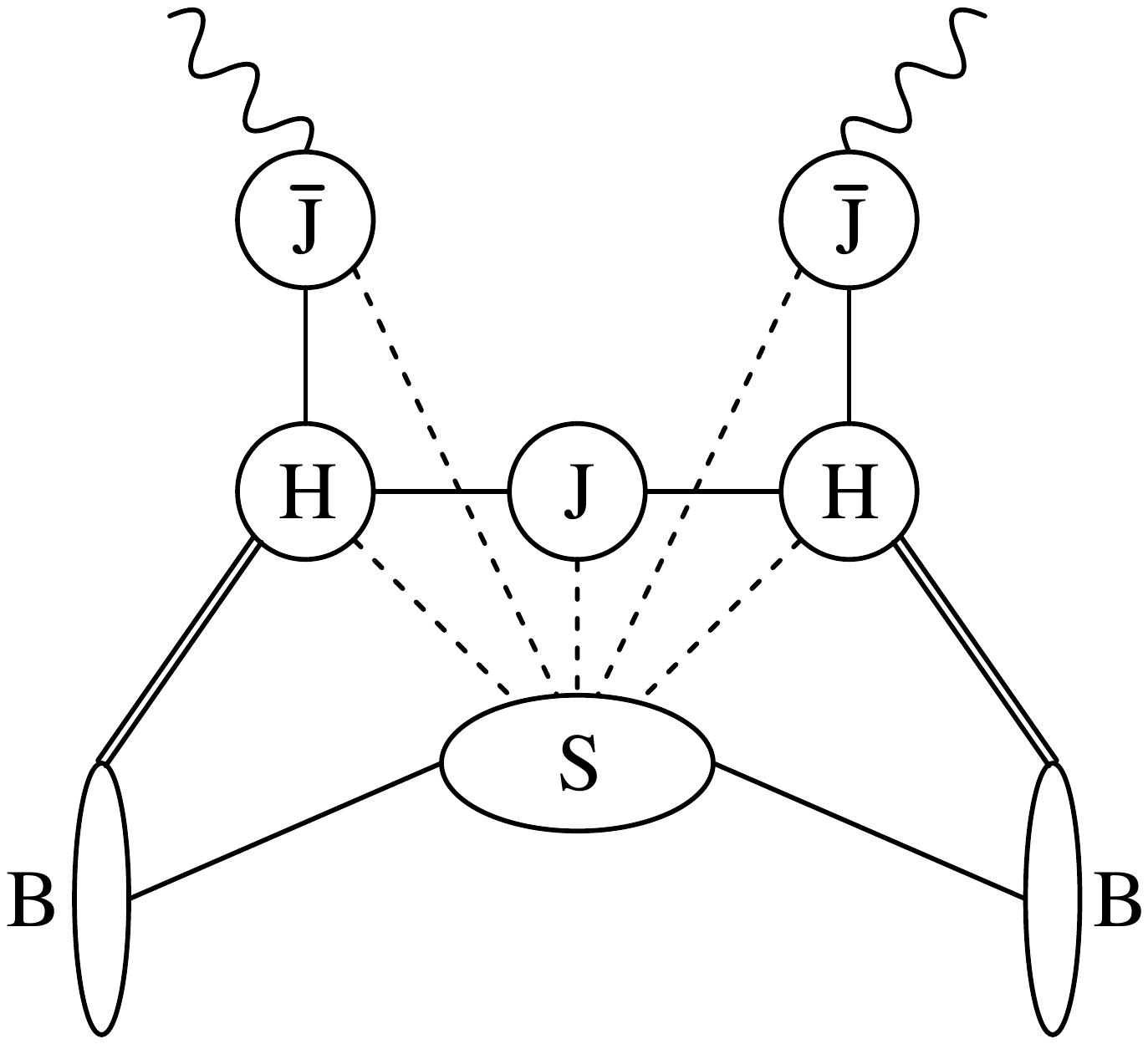,width=5.9cm}
{\caption{\label{Fig:theorem}
Graphical illustration of the three terms in the QCD factorization theorem (\ref{fact2}). The dashed lines represent soft interactions, which must be power expanded and factored off the remaining building blocks to derive factorization.}}
\end{center}
\end{figure}

We therefore describe the hadronic effects with SCET, corresponding to an expansion of the forward scattering amplitude in non-local operator matrix elements. 
One derives a  factorization formula for the considered process, in complete analogy to the radiative decay in \cite{Benzke:2010js}
\begin{align}\label{fact2}
   &d\Gamma(\bar B\to X_s\ell^+ \ell^-)
   = \sum_{n=0}^\infty\,\frac{1}{m_b^n}\, \sum_i\,H_i^{(n)} J_i^{(n)}\otimes S_i^{(n)} \nonumber \\
   &\qquad + \sum_{n=1}^\infty\,\frac{1}{m_b^n}\,\bigg[ \sum_i\,H_i^{(n)} J_i^{(n)}\otimes S_i^{(n)}\otimes\bar J_i^{(n)}
    + \sum_i\,H_i^{(n)} J_i^{(n)}\otimes S_i^{(n)} \otimes\bar J_i^{(n)}\otimes\bar J_i^{(n)} \bigg] \,. 
\end{align}
The formula contains the so-called direct contributions in the first line, while the second line describes the resolved contributions which occur first only at the order $1/m_b$ in the heavy-quark expansion.  
Fig.~\ref{Fig:theorem}  shows a graphical illustration of the three terms in the factorization theorem in the shape function region. 
Here $H_i^{(n)}$ are the hard functions describing physics at the high scale $m_b$, $J_i^{(n)}$ are so-called jet functions characterizing the physics of the hadronic final state $X_s$ with the invariant mass in the range described above.
The hadronic physics associated with  the scale $\Lambda_\text{QCD}$ is parametrized by the soft functions $S_i^{(n)}$. Similarly to the radiative decay investigated in Ref.~\cite{Benzke:2010js}, we have in addition resolved virtual-photon  contributions in the second line, whose effects are
described by new jet functions $\bar J_i^{(n)}$. This occurs due to the coupling of virtual photons with virtualities of order $\sqrt{m_b \Lambda_\text{QCD}}$ to light partons instead of the weak vertex directly. Consequently they probe the hadronic substructure at this scale.
Resolved effects  may occur as a single or double ``resolved'' contribution due to interference of the various operators, which also have the ``direct virtual-photon'' contribution.
Finally the soft or shape functions are defined in terms of forward matrix elements of non-local heavy-quark effective theory (HQET) operators. This limited set of shape functions can not be calculated perturbatively, yet
this allows a systematic analysis of hadronic effects in this decay mode.
We imply the convolution of the soft and jet function due to the occurrence of common variables with the symbol $\otimes$. Finally, we note that --  as already discussed in Ref.\cite{Benzke:2010js}  --   there is not a complete proof of this factorization formula. There is one case in which there is a UV divergent convolution integral within the resolved contribution.       
The contribution from ${\cal O}_8 - {\cal O}_8$ possesses an ultraviolet divergence, which cancels the $\mu$-dependence of the corresponding subleading jet function. This cancellation is expected and needed. However, a proper factorization of the anti-jet functions is needed to have a consistent description. Thus, this issue has been fixed by considering the convolution of the two anti-jet functions with the soft-function. The limit of the DimReg parameter $\epsilon$ needs to be taken after the convolution has been performed  in order to obtain the proper factorization result, but this is contradictory to the assumption given in the factorization formula.~\footnote{We note that there are also divergent convolution integrals in SCET in power-suppressed contributions to hadronic $B$ meson decays. The important difference to our present case is that these divergences have an IR-origin.}

Within this context, we consider only the low $q^2$ region. In this  region, obeying the invariant mass constraint, the only sensible power-counting - as shown above - is to assume $q$ scales as an anti-hard-collinear momentum, while $P_X$ as a hard-collinear momentum just as in the radiative decay. In this  sense, at least one of the leptons has to be anti-hard-collinear, while the other may be soft. 
In our effective theory, we have, besides the initial heavy quark, active hard-collinear  and anti-hard-collinear fermions, whose fields scale in $x$-space as
\begin{equation}
  \xi_\text{hc} = W_{\bar n}^\dagger \xi_n \sim \sqrt{\lambda}\,, \qquad \xi_{\overline{\text{hc}}} = W_{ n}^\dagger \xi_{\bar n} \sim \sqrt{\lambda}\,.
\end{equation}
These two-component spinor fields obey the projector identities $P_n \xi_n = \xi_n$, $P_{\bar n} \xi_n = 0$ (and $n \leftrightarrow \bar n$) with $P_n = \frac{\slashed{n} \slashed{\bar n}}{4}$ and  $P_{\bar n} = \frac{\slashed{\bar n} \slashed{ n}}{4}$. The quantities $W_n\,,W_{\bar n}$ are the familiar (anti)-hard collinear Wilson lines in SCET  that render the Lagrangian gauge invariant. The soft and heavy quark fields scale as $h, q \sim \lambda^{3/2}$. The $b$-quark is described in terms of a HQET field and its velocity is given by $v^\mu = 1/2 (n^\mu + \bar{n}^\mu)$ and to leading order the $b$-field satisfies $\slashed{v} b = b$.

The projections of the gauge fields onto the components scale the same as the corresponding momentum components
\begin{equation}
    \mathcal{A}^\mu_\text{hc} = W_{\bar n}^\dagger ( i D^\mu_\text{hc} W_{\bar n}) \sim (\lambda, 0, \sqrt{\lambda})\,\qquad\mathcal{A}^\mu_{\overline{\text{hc}}} = W_{n}^\dagger ( i D^\mu_{\overline{\text{hc} }} W_{n}) \sim (0, \lambda, \sqrt{\lambda})\,.
\end{equation}
Using this scaling, we can match the operators \eqref{eq:op_basis} onto the corresponding SCET operators and order them according to the scaling parameter $\lambda$. The relevant SCET Lagrangian for hard-collinear and soft fields (for anti-hard we need to replace $n \leftrightarrow \bar n$) obtained by the matching from the simple QCD (QED) Lagrangian is given by \cite{Beneke:2002ph,Beneke:2002ni}
\begin{equation} \label{eq:scet_lagrange}
{\cal L} = \bar{\xi}_n \left(i n\cdot D + i \slashed{D}_{\perp hc} \frac{1}{i \bar n\cdot D_{hc}}\, i\slashed{D}_{\perp hc} \right)
 \frac{\slashed{\bar n}}{2} \, \xi_n +   \bar{q}\, i \slashed{D}_{\rm s}(x) q + {\cal L}_{\xi q}^{(1)} ,
\end{equation}
where the superscript denotes the suppression in powers of $\sqrt{\lambda}$. The terms are explicitly given by
\begin{align}
    {\cal L}^{(1)}_{\xi q} &=  \bar q \,W_{\bar n}^\dagger i\slashed{D}_{\perp hc} \,\xi_n - 
    \bar{\xi}_n \,i\overleftarrow{\slashed{D}}_{\perp hc} W_{\bar n}\, q \, . \label{eq:scet_colsoft_sl}
\end{align}
In order to describe the process in question, we need to combine $\text{QCD}\otimes\text{QED}$ in terms of SCET. Kinematically we are in the situation where we need to describe the hadronic part in terms of SCET for a proper and consistent description, but also as far as QED is concerned, we have to describe these fields in terms of an SCET-like theory.
Thus, we investigate the matching of ${\cal O}_7$ onto SCET fields, where we consider the (virtual) photon to be power-counted as well.

The electromagnetic dipole operator is then written as
 \begin{equation}
   {\cal O}_{7\gamma}= -\frac{e}{8\pi^2}\,m_b\,
    \bar s\sigma_{\mu\nu}(1+\gamma_5) F^{\mu\nu} b 
\,.
\end{equation}
Suppressing the $-\frac{e m_b}{4\pi^2}\,e^{-im_b\,v\cdot x}$ factor and following the notation of \cite{Beneke:2002ph} ${\cal O}_{7\gamma}$ is matched onto the leading operator with $\mathcal{A}^\text{em}$ being the Wilson line dressed gauge-invariant photon field
\begin{equation}\label{eq:Q7_A}
{\cal O}_{7\gamma A}^{(0)} = \bar{\xi}_\text{hc}\,\frac{\bar{\slashed{n}}}{2}\, [i n\cdot \partial \slashed{\mathcal{A}}^\text{em}_{\perp}]\,(1+\gamma_5)  h  \,. 
\end{equation}
We count the photon field as $ (n\cdot \mathcal{A}^\text{em}, \bar n\cdot \mathcal{A}^\text{em}, \mathcal{A}^\text{em}_\perp) \sim ( 0, \lambda , \sqrt{\lambda})$, where $n\cdot \mathcal{A}^\text{em} = 0$ follows from gauge invariance, despite of being off-shell.

We need to contract the photon from this operator with the QED Lagrangian in order to convert this virtual photon into a lepton-antilepton pair.  Note that the contribution of ${\cal O}_7$ scales as $\lambda^\frac{5}{2}$. 
The conversion of the virtual photon into hard-collinear leptons introduces no further suppression.
For the semi-leptonic operators, the matching leads to the following SCET operators
\begin{align}
    {\cal O}_9 &= \frac{\alpha}{2\pi} (\bar sb)_{V-A} (\bar ll)_{V} &\quad &\rightarrow &\quad {\cal O}_9^{(1)} &= \frac{\alpha}{2\pi} ( \bar{\xi}_\text{hc}^s  [1+\gamma^5 ] h) ( \bar{\xi}_\text{hc}^\ell \frac{\slashed{n}}{2}  \xi_\text{hc}^\ell)\label{eqn:Q9scet} \\
    {\cal O}_{10} &= \frac{\alpha}{2\pi} (\bar sb)_{V-A} (\bar ll)_{A} &\quad &\rightarrow &\quad   {\cal O}_{10}^{(1)} &=\frac{\alpha}{2\pi} ( \bar{\xi}_\text{hc}^s [1+\gamma^5 ] h) ( \bar{\xi}_\text{hc}^\ell \frac{\slashed{n}}{2}  \gamma^5 \xi_\text{hc}^\ell)\label{eqn:Q10scet}\,.
\end{align}
Both operators scale as $\lambda^{\frac12 +\frac32 +2 \frac12} = \lambda^3$, which is suppressed by $\lambda^\frac12$ against the contribution from ${\cal O}_7$. Note that this changes in the high $q^2$ region as in this case the leptons are hard and do not add a power suppression. 

{Thus, according to the power counting in the low $q^2$ region, the leading order reference is given by the direct ${\cal O}_7 - {\cal O}_7$ contribution at the order of $\lambda^5$. If one takes  into account all contributions up to order $1/m_b$ corrections, i.e. terms up to $\lambda^6$, corresponding to ${O}(\lambda)$ corrections to  the leading direct contribution, then within the direct contributions one  has to include  only the leading part of ${\cal O}_{9,10} - {\cal O}_{9,10}$, but the subleading part of ${\cal O}_7 - {\cal O}_7$. This includes  subleading soft and jet functions and the resolved contributions due to  interference with other operators.} 

In this paper, we  calculate  the resolved contributions, which we consider to order $1/m_b$. For this, we need to compute the resolved contributions from ${\cal O}_1 - {\cal O}_7$, ${\cal O}_7 - {\cal O}_8$ and ${\cal O}_8 - {\cal O}_8$ 
{ as in the $\bar B\to X_s\gamma$. They appear at the same order in the power counting in   $\bar B\rightarrow X_s \ell^+ \ell^-$ , since the conversion of the photon into the hard collinear leptons is not power suppressed. Are there additional contributions? Indeed, the virtual photon could  give rise to additional quantities in the operator matching, which where zero in the real case. In particular, subleading operators might contain factors of $\bar n\cdot q$ and $\bar n\cdot{\cal A}^\text{em}$. However, these operators contain the photon field directly (i.e. they do not couple to the photon via a Lagrangian insertion), and therefore do not give rise to resolved contributions. Also, there are no additional operators at leading power  that contain these factors.}

The usual observables can be obtained from the triple differential rate in the form
\begin{equation}
    \frac{\text{d}^2 \Gamma}{ \text{d}q^2 \text{d} z} = \frac38 \left[(1 + z^2) H_T (q^2) + 2 (1 - z^2) H_L(q^2) + 2 z H_A(q^2) \right] \label{eq:DiffRate}
\end{equation}
as given in \cite{Huber:2015sra}, and we will calculate the corrections to the structure functions $H_i$ below.

\section{Obtaining the Fully Differential Decay Rate}
\label{sec:differential}

The differential rate is obtained by calculating the restricted discontinuity 
\begin{align}\label{eq:Discontinuity}
   d\Gamma(\bar B\to X_s\ell^+ \ell^-) \propto \text{Disc}_\text{\,restr.}\,
   \Big[ i\int d^4 x\,\langle\bar B| {\cal H}_\text{eff}^\dagger(x)\,
   {\cal H}_\text{eff}(0) |\bar B\rangle \Big] 
\,,
\end{align}
where the restriction implies that only cuts that contain the appropriate final states are taken into account.
At first order in the electromagnetic coupling the resulting expression can be decomposed into a hadronic and a leptonic tensor, $W^{\alpha \beta}$ and $L_{\alpha \beta}$ respectively
\begin{align}
   d\Gamma(\bar B\to X_s\ell^+ \ell^-) = d\Pi^\text{Lept}\, L_{\alpha \beta}(p_{\ell^+},p_{\ell^-})\, W^{\alpha \beta}(v,p_{\ell^+}+p_{\ell^-}) \,,
\end{align}
with the leptonic phase space indicated by $d\Pi^\text{Lept}$.
The hadronic tensor $W^{\alpha \beta}$ contains the integration over the final state hadronic momentum and the total momentum conservation in its definition
\begin{equation}
  W_{\alpha \beta} = \sum_{X_s}\int\frac{\text{d}^3 p_{X_s} }{(2\pi)^3 2E_{X_s}}  \frac{1}{2 M_B}\langle B | {\cal O}^{\dagger,\text{had}}_\beta | X_s \rangle\langle X_s|{\cal O}^\text{had}_\alpha | B \rangle  (2\pi)^4 \delta^{(4)} (P_B - p_{X_s} - p_{\ell^+}-p_{\ell^-} )\,.
\end{equation}
with the Fourier transformed operators ${\cal O}^\text{had}_\alpha$. This explicitly contains the on-shell condition. 
For the leptonic tensor we have to distinguish between the contribution from the QED current insertion, and the direct contributions from ${\cal O}_{9,10}$, with the former defined as including the virtual photon propagator
\begin{align}
L_{\alpha \beta}^\text{QED} = -\left (\frac{-i}{q^2}\right)^2 \big(- i e  \big)^2\,
  \text{tr}\big(\slashed{p}_{\ell^+}\gamma_\alpha\,\slashed{p}_{\ell^-}\gamma_\beta\big)\,.
\end{align}
As will be shown below, for the current insertions only terms containing perpendicular components survive the contraction with the hadronic tensor to the first order. For the semi-leptonic contributions on the other hand the leptonic tensor is contracted with $n^\alpha n^\beta$, which can be seen from equations (\ref{eqn:Q9scet}) and (\ref{eqn:Q10scet}). But as explained above, there are no resolved contributions with the semi-leptonic operators to the first order in $1/m_b$. Thus, we can  restrict ourselves to the insertion of a QED current in the following.

Below, the resolved $1 / m_b$ corrections to this hadronic tensor are calculated within the framework of SCET.  Any desired distribution can then be recovered by performing the phase-space integration over the lepton momenta outlined below for our numerical study. 

For an unpolarized three body decay we have two degrees of freedom. Remember that the hadronic on-shell condition leads to a delta distribution, respectively its derivative for power corrections, in the hadronic tensor. This condition is implicitly contained in the non-local matrix element, and therefore we can have at most a triple differential rate from the phase space, where this on-shell condition still needs to be evaluated. It is convenient to use the following three kinematic variables as it was already indicated in Eq.~\eqref{eq:DiffRate}, 
\begin{align}
   v\cdot q\,\,;
   q^2\,;\,
    z &= \cos \theta = \frac{v\cdot p_{\ell^+} - v\cdot p_{\ell^-} }{ \sqrt{(v\cdot p_{\ell^-} + v\cdot p_{\ell^+} )^2 - q^2} \sqrt{1- 4\frac{ m_\ell^2 }{q^2}}}\,, \label{eq:LeptonAngle}
\end{align}
where  $q =  p_{\ell^+} +  p_{\ell^-}$, $v = 1/2 (n +\bar n)$, and  $z$ is defined as the angle of the positively charged anti-lepton with the flight direction of the
$B$-meson in the rest frame of the lepton-antilepton system ($\vec q = 0$). We keep the leptons massless in the following discussion. Then the structure functions in Eq.~\eqref{eq:DiffRate} can easily be identified.
In this notation it is obvious  that $z$ is a Lorentz scalar, and in the $B$-rest frame $v\cdot p_{\ell_\pm} = E_{\ell_\pm}$.

We derive the phase-space result in full QED kinematics. It can be shown  that expanding this calculation to the leading order in $\lambda$ is equal to the result calculated directly in leading order SCET.
Furthermore it is easy to verify  that the leptonic part $I_{\alpha \beta}(v,q,z)$ defined in the contraction 
\begin{align}
&\int d\Pi^\text{Lept}\, L_{\alpha \beta}^\text{QED}(p_{\ell^+},p_{\ell^-})\, W^{\alpha \beta}(v,p_{\ell^+}+p_{\ell^-})\nonumber\\
=\,&\int d\Pi^\text{Lept}\,\frac{d^4 q}{(2\pi)^4}\,(2\pi)^4 \delta^{(4)} (q-p_+-p_-)\,dz\,\delta\left(z - \frac{v\cdot p_{\ell^+} - v\cdot p_{\ell^-} }{ \sqrt{(v\cdot q)^2 - q^2} } \right)\nonumber\\
& \times\frac{-e^2}{(q^2)^2} \,
  \text{tr}\big(\slashed{p}_{\ell^+}\gamma_\alpha\,\slashed{p}_{\ell^-}\gamma_\beta\big)\, W^{\alpha \beta}(v,q)\nonumber\\
\equiv\,&\int dv\cdot q\,dq^2\,dz\,\frac{\sqrt{v\cdot q^2 -q^2}}{(2\pi)^3}\,\frac{4\pi\alpha}{(q^2)^2}\big(- I_{\alpha \beta}(v,q,z) \big)\,W^{\alpha \beta}(v,q)\label{Eq:WLcont}
\end{align}
is transforming as a tensor under Lorentz transformations. Here, we have explicitly included the dependence on the angle $z$. The only invariants  which occur in the integrand are $v\cdot p_{\ell^-}$ and $q  \cdot p_{\ell^-}$. Therefore, using current conservation $q_\mu L^{\mu \nu} = 0 = q_\nu L^{\mu \nu}$ for massless leptons, we can decompose $I_{\alpha \beta}(v,q,z)$ as
\begin{align}
    I^{\alpha \beta} (v, q, z) = \phantom{+}& I_1 (v\cdot q, q^2, z) \left( - g^{\alpha \beta} + \frac{q^\alpha q^\beta}{q^2} \right) \nonumber \\
    +&I_2 (v\cdot q, q^2, z) \left(v^\alpha v^\beta + \frac{q^\alpha q^\beta (v\cdot q)^2}{q^4} - \frac{(v^\alpha q^\beta +v^\beta q^\alpha)(v\cdot q)}{q^2}\right)  \nonumber \\
    +&I_3 (v\cdot q, q^2, z) i \epsilon^{\alpha \beta \rho \sigma} v_\rho q_\sigma \,.\label{eqn:Idecomp}
\end{align}
Note that for the same reasons  we may decompose the hadronic tensor $W^{\alpha \beta} (v,q)$ in a similar way, as it depends on $v^\mu$ and $q^\mu$, only.

In the case relevant for the resolved contribution we have to explicitly compute this decomposition for the insertion of a QED current. Then the leptonic structure functions are given by
\begin{align}
    I_1 (v\cdot q, q^2, z) &= -\frac{q^2}{16 \pi} (1+z^2)\\
    I_2 (v\cdot q, q^2, z) &= -\frac{q^2}{16 \pi} \frac{q^2}{(v\cdot q)^2-q^2} (1-3 z^2)\\
    I_3 (v\cdot q, q^2, z) &= 0\,.
\end{align}
The absence of a linear component in $z$ shows  that there exists no resolved contribution to the forward-backward asymmetry at this order. However, this result is already anticipated as neither ${\cal O}_9$ nor ${\cal O}_{10}$ contribute for resolved corrections at this order.

Expanding this result to order ${O}(\lambda)$, where we have to take into account that $q_\perp=0$ and that the open indices couple to a virtual photon field scaling as anti-hard-collinear, we obtain
\begin{equation}
 I^{\alpha \beta} (v, q, z) = - g_\perp^{\alpha \beta} \frac{n\cdot q\, \bar{n} \cdot q}{16 \pi} (1+z^2) + {O}(\lambda)\,.
\end{equation}
In this sense, the Dirac structure reduces to the on-shell photon case. Combining this expanded result with the phase-space integration in Eq.~\eqref{Eq:WLcont}, we obtain
\begin{align}
&d\Pi^\text{Lept}\, L_{\alpha \beta}^\text{QED}(p_{\ell^+},p_{\ell^-})\, W^{\alpha \beta}(v,p_{\ell^+}+p_{\ell^-})\nonumber\\
= &\,dv\cdot q\,dq^2\,dz\, \frac{\alpha}{32\pi^3} (1+z^2)\,\frac{\sqrt{v{\cdot}q^2 - q^2}}{q^2} \,g_{\perp,\alpha \beta}\,W^{\alpha \beta}(v,q)\nonumber\\
\equiv &\, d\Lambda_{\alpha\beta}\,W^{\alpha \beta}(v,q)\,,
\end{align}
where we have defined the abbreviation $d\Lambda_{\alpha\beta}$ for later convenience. The transition to light-cone coordinates is easily obtained by using 
\begin{align}
    n \cdot q &= v\cdot q + \sqrt{v{\cdot}q^2 - q^2}\\
    \bar{n} \cdot q &= v\cdot q - \sqrt{v{\cdot}q^2 - q^2}\,.
\end{align}
for an anti-hard-collinear momentum $q$.  Neglecting $\lambda$ corrections it is easy to calculate
\begin{equation}
    \text{d}v {\cdot} q \text{d} q^2 = \frac{n \cdot q}{2} \text{d}n{\cdot} q \text{d} \bar{n}{\cdot} q\,.
\end{equation}
where we have approximated $\sqrt{(v\cdot q)^2 - q^2} \approx \frac12 n\cdot q$.
Furthermore we find that in comparison with Eq.~\eqref{eq:DiffRate}, the only structure function that gets corrections of this type to the considered order is $H_T (q^2)$, while $H_A(q^2)$ and $H_L(q^2)$ do not. 
Thus, we find 
\begin{align}
d\Lambda_{\alpha\beta} =  
dn\cdot q\,d\bar n\cdot q\,dz \frac{\alpha}{128\pi^3} (1+z^2)\,\frac{n\cdot q}{\bar n\cdot q} \,g_{\perp,\alpha \beta}\,. 
\end{align}
With the appropriate replacement derived above we can therefore transit between the two differential rates, where we have to obey the power-counting in replacing the variables, by
\begin{equation}
 \frac{ \text{d}^3 \Gamma}{\text{d}v {\cdot} q \text{d} q^2\,\text{d} z} = \frac{4 \bar n\cdot q}{n\cdot q} \frac{\sqrt{v{\cdot}q^2 - q^2}}{q^2} \frac{ \text{d}^3 \Gamma}{\text{d}n{\cdot}q\,\text{d}\bar n {\cdot}q\,\text{d} z}
\end{equation}
Finally, we can compare our results to the already known results of $B\rightarrow X_s \gamma$. This can be done by recomputing the phase-space and setting $\bar n \cdot q = 0$
\begin{equation}
d\Gamma(\bar B\to X_s\gamma) = dE_\gamma\, \frac{n\cdot q}{8\pi^2} \,g_{\perp,\alpha \beta} \,W^{\alpha \beta}(v,q)\label{Eq:PSgamma}\,.
\end{equation}
This corresponds to
\begin{align}
      \frac{4\pi}{\alpha (1+z^2)}\frac{n\cdot q\,\,q^2}{\sqrt{v{\cdot}q^2 - q^2}} \frac{ \text{d}^3 \Gamma}{\text{d}v {\cdot} q \text{d} q^2\,\text{d} z} \bigg|_{\bar n\cdot q \rightarrow 0} &\rightarrow \frac{ \text{d} \Gamma}{\text{d}E_\gamma} \nonumber \\
     \frac{16\pi}{\alpha (1+z^2)} \bar n\cdot q \frac{ \text{d}^3 \Gamma}{\text{d}n{\cdot}q\,\text{d}\bar n {\cdot}q\,\text{d} z} \bigg|_{\bar n\cdot q \rightarrow 0} &\rightarrow \frac{ \text{d} \Gamma}{\text{d}E_\gamma}\,.
\end{align}

\newpage

\section{\texorpdfstring{Explicit calculation of the  ${\cal O}_1 - {\cal O}_{7\gamma}$ contribution}{Explicit calculation of the O1-O7 contribution}}
\label{sec:example}

\begin{figure}
\begin{center} 
\epsfig{file=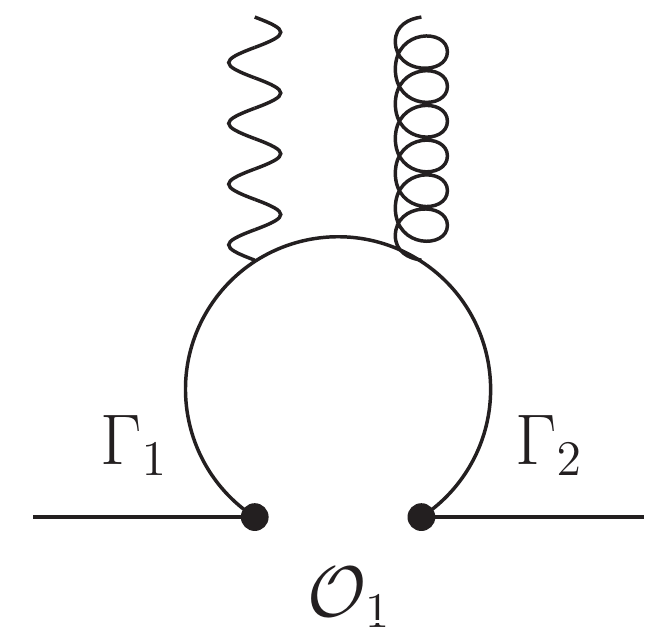,width=5cm}\hspace{1cm}\epsfig{file=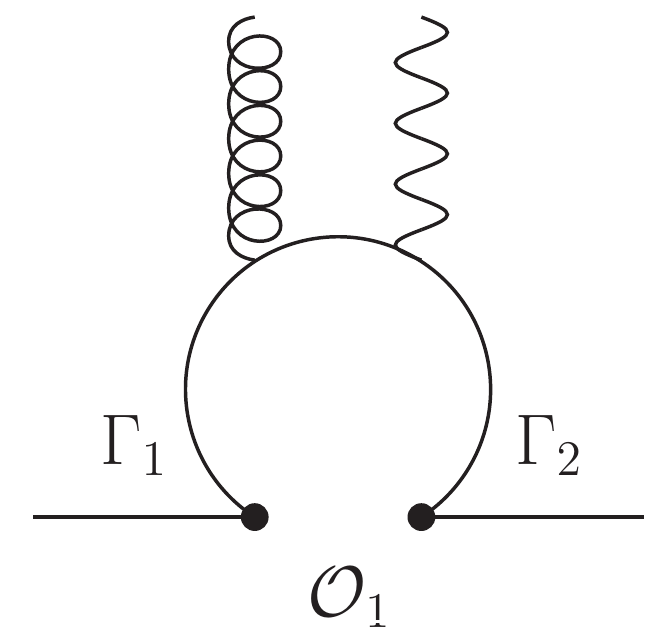,width=5cm}
{\caption{\label{Fig:Q1loop}
Graphical illustration of the leading charm-quark loop contribution with the emission of an off-shell photon  and a soft gluon  induced by the operator }}
\end{center}
\end{figure}

For the explicit calculation of this resolved contribution  we need to derive the expression for the loop with the emission of an anti-hard-collinear virtual photon and a soft gluon. We define (see Fig.~\ref{Fig:Q1loop})

\begin{equation}
{\cal A} = \frac{e q_q}{4\pi}\,\frac{g}{4\pi}\,\bar s\Gamma_2\,A\,\Gamma_1 b\,.
\end{equation}\\
Considering only those contributions that do not vanish between the Dirac structures $\Gamma_2\otimes\Gamma_1=\gamma^\mu(1-\gamma_5)\otimes\gamma_\mu(1-\gamma_5)$ the leading charm-quark loop contribution with the emission of an off-shell photon $q$ and a soft gluon $l_1$ is given in gauge invariant form by
\begin{align}
A = 
\frac{i \gamma_{\beta } \gamma^5}{2\left(l_1\cdot q\right){}^2}
&\bigg[ \left(F_{\mu \alpha} \tilde{G}^{\mu \beta}+G_{\mu \alpha}\tilde{F}^{\mu \beta }\right) (q+l_1)^{\alpha }
\Big \lbrace 
  (q+l_1)^2\left(1-F\left(\frac{m_c^2}{(q+l_1)^2}\right)\right) \nonumber\\
&- q^2\left(1-F\left(\frac{m_c^2}{q^2}\right)\right) - q^2\left(G\left(\frac{m_c^2}{(q+l_1)^2}\right) - G\left(\frac{m_c^2}{q^2}\right)\right)
\Big \rbrace
\nonumber \\
&-  F_{\mu \alpha } \tilde {G}^{\mu \beta } q^{\alpha } \Big \lbrace 
  (q+l_1)^2\left(1-F\left(\frac{m_c^2}{(q+l_1)^2}\right)\right) \nonumber\\
&- q^2\left(1-F\left(\frac{m_c^2}{q^2}\right)\right) - (q+l_1)^2\left(G\left(\frac{m_c^2}{(q+l_1)^2}\right) - G\left(\frac{m_c^2}{q^2}\right)\right)
\Big\rbrace \bigg]\,,
\end{align}
where we are using the convention  
\begin{equation}
\tilde F^{\mu\nu} = -\frac{1}{2}\epsilon^{\mu\nu\alpha\beta}F_{\alpha\beta}\quad(\epsilon^{0123}=-1)\,,
\end{equation}
and have defined the penguin functions
\begin{align}
F(x)&=4x\arctan^2\frac{1}{\sqrt{4x-1}}\,, \label{FF}\\
G(x)&=2\sqrt{4x-1}\arctan\frac{1}{\sqrt{4x-1}}-2 \,.
\end{align}
For a real photon $q^2=0$ and $q^\alpha F_{\alpha\beta}=0$ the above expression reduces to
\begin{align}
A = 
\frac{i \gamma_{\beta } \gamma^5}{2\left(l_1\cdot q\right){}^2}
\bigg[ \left(F_{\mu \alpha} \tilde{G}^{\mu \beta}+G_{\mu \alpha}\tilde{F}^{\mu \beta }\right) (q+l_1)^{\alpha }
\Big \lbrace 
  2\,l_1\cdot q\left(1-F\left(\frac{m_c^2}{2\,l_1\cdot q}\right)\right)\Big\rbrace \bigg]\,,
\end{align}
and we reproduce the result from $B \rightarrow X_s \gamma$. Note, that in the soft limit, where also $l_1\cdot q\to 0$ the product of the prefactor $1/(l_1\cdot q)^2$ with the specific combination of the penguin functions given above remains finite. As far as the field-strength tensors are concerned, the leading power is given by 
\begin{align}
 &(q+l_1)^\alpha \gamma_\beta \gamma^5 (G_{\mu \alpha} \tilde F^{\mu \beta} )\nonumber\\
=&\,\frac{1}{4}\,(n\cdot q)^2 i\epsilon^{\beta\sigma\mu\rho}\bar n_\rho\, \bar n^\alpha G_{\perp\alpha\mu}\,\epsilon_{\perp\sigma}^{(\gamma)*}
+{\cal O}(\lambda^3)\,,
\end{align}
where the polarization vector $\epsilon^{(\gamma)}$ represents an off-shell photon, which gives rise to the anti-hard-collinear propagator, when contracted with the QED current. Calculating the interference with the operator ${\cal O}_{7\gamma}$ we obtain the differential rate as
\begin{align}
d\Gamma_{1 7} =& \frac{1}{m_b}\,\text{Re}\Big[\hat\Gamma_{1 7}\frac{-\lambda_t^*\lambda_c}{|\lambda_t|^2}\Big]\,
d\Lambda_{\alpha\beta}\, e_c\, (n\cdot q)^2\,
\text{Re}\int d\omega \delta( \omega + m_b - n\cdot q)\int d\omega_1 \frac{1}{\omega_1 + i \epsilon}\nonumber\\
&\times\frac{1}{\omega_1}\left[
(\nb\cdot q+\omega_1)\left(1-F\left(\frac{m_c^2}{n\cdot q(\nb\cdot q+\omega_1)}\right)\right)
-\nb\cdot q \left(1-F\left(\frac{m_c^2}{n\cdot q\nb\cdot q}\right)\right) \right.\nn\\
&\left.-\nb\cdot q \left( G\left(\frac{m_c^2}{n\cdot q(\nb\cdot q+\omega_1)}\right) - G\left(\frac{m_c^2}{n\cdot q\nb\cdot q}\right)\right)\right]\nonumber\\
&\times\int \!\frac{dt}{2\pi} e^{-i \omega t}\!\!\int \!\frac{dr}{2\pi} e^{-i \omega_1 r}
\frac{\langle B | \bar h(nt) \slashed{\bar n} [1+\gamma^5]\frac{i}{2} [\gamma^\mu_\perp, \gamma^\beta_\perp]\gamma_\perp^\alpha \bar{n}^\kappa g G_{\mu \kappa}(\bar n r)  h(0) | B \rangle}{2M_B}
\end{align}
where we have defined the shorthand notation
\begin{equation}
\hat\Gamma_{i j} = \frac{G_F^2\alpha m_b^2}{4\pi^2}\, C_i C_j^*\,|\lambda_t|^2\,,
\end{equation}
and the $i\epsilon$ prescription may be dropped if we assume the soft function is well behaved in the limit $\omega_1\to 0$. The result obviously reproduces the known structure function result in the limit of a real photon. 
For this we have to replace the leptonic tensor by $- g_{\kappa \sigma}$, the photon energy by $n\cdot q = 2 E_\gamma$ and $\bar n \cdot q = 0$. We then obtain for the contraction of the matrix element
\begin{align}
  g_{\alpha \beta}[\gamma^\mu_\perp, \gamma^\beta_\perp]\gamma_\perp^\alpha
  = 2 \gamma^\mu_\perp \,,
\end{align}
which exactly reproduces the soft function in the radiative decay. 
{The same is true for the semi-leptonic decay. Due to $q_\perp=0$ the only remaining term of the decomposition of the leptonic tensor in (\ref{eqn:Idecomp}) is again $g_\perp^{\alpha\beta}$ and the Dirac structure in the shape function again reduces to the radiative case.}
Hence, no new structure function is involved to this order in the power-counting.

\newpage

\section{\texorpdfstring{Results to first order in $1/m_b$}{Results to first order in 1/mb}}
\label{sec:contributions}

Using standard relations explained in section \ref{sec:differential}, we automatically achieve the decomposition of the hadronic tensor into Lorentz structure functions. Below we have listed the results for the resolved contributions at order $\lambda$ for the hadronic tensor. The smooth limit $q^2\rightarrow 0$ reproduces the known results from Ref.~\cite{Benzke:2010js}. In the following we state our  results for the CP-averaged rate, i.e. the result
factorizes into the real part of the strong matrix element and the weak prefactors. 
We have  three resolved operator combinations  to order $1/m_b$. Namely ${\cal O}_{7\gamma} - {\cal O}_{8g}$,\, ${\cal O}_{8g} - {\cal O}_{8g}$, and ${\cal O}_1 - {\cal O}_{7\gamma}$.

\begin{figure*}[t!]
\begin{center}
\includegraphics[scale=0.4]{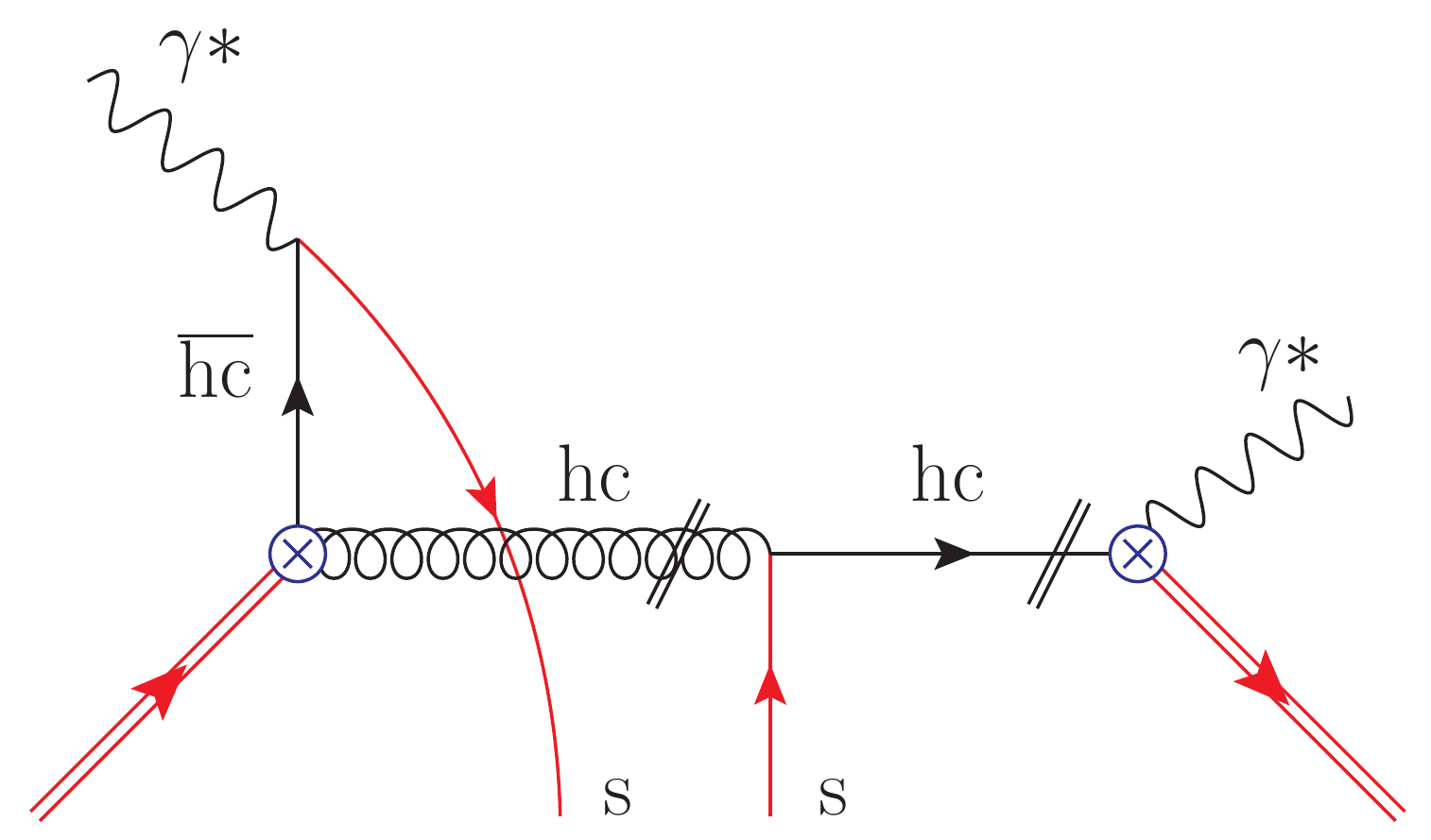}\includegraphics[scale=0.4]{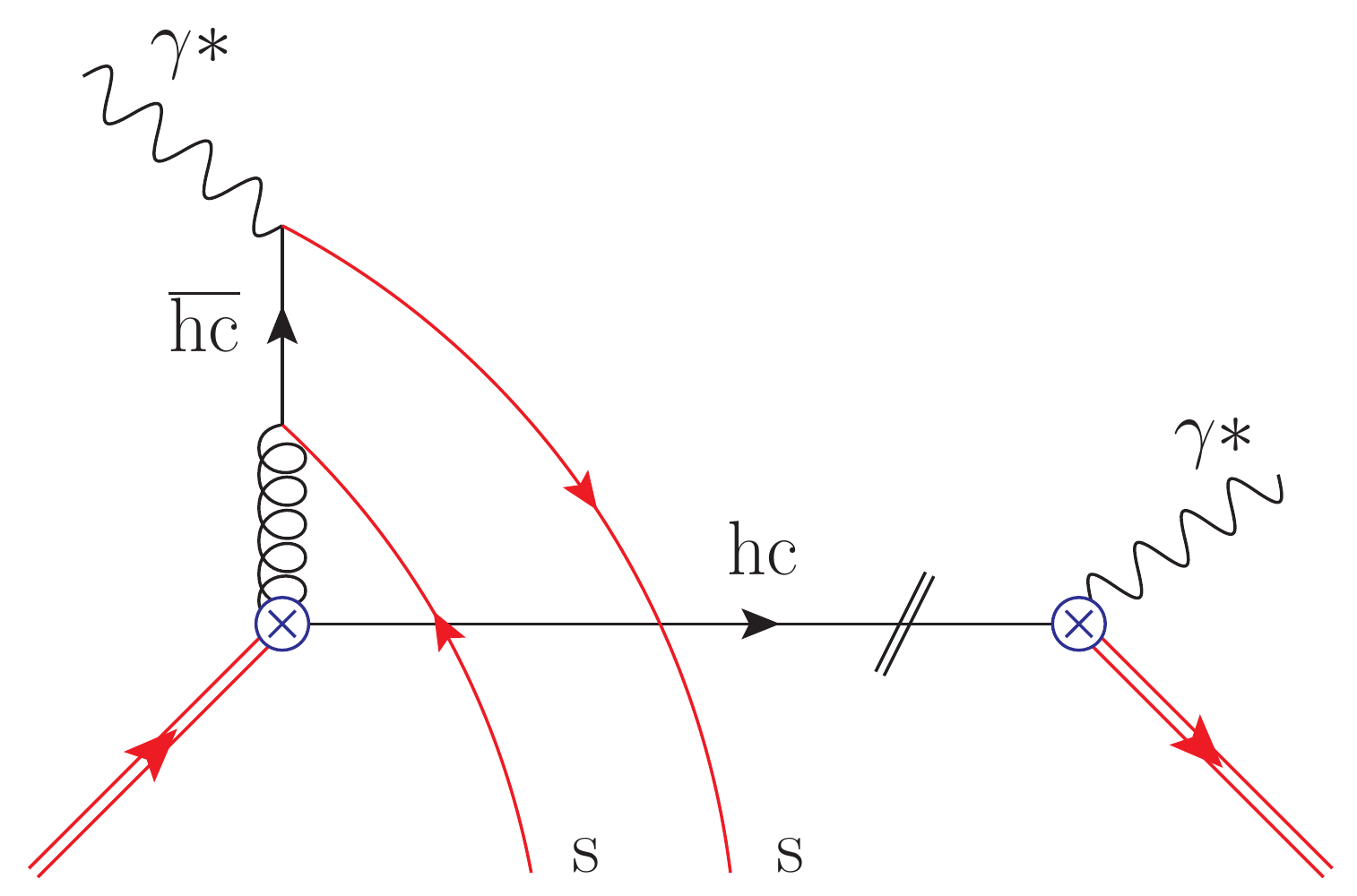}
    \caption{Three cut diagrams arising from the matching of the  ${\cal O}_{7\gamma} - {\cal O}_{8g}$ contribution onto SCET. Red indicates soft fields, black (anti-) hard-collinear fields. Hard fields are already integrated out. The left diagram with a hard-collinear gluon allows for two different cuts, while the diagram with the anti-hard-collinear gluon allows for one cut only.}  \label{Fig:diagrams0708}
\end{center}
\end{figure*}

\begin{figure*}[t!]
\begin{center}
\includegraphics[scale=0.4]{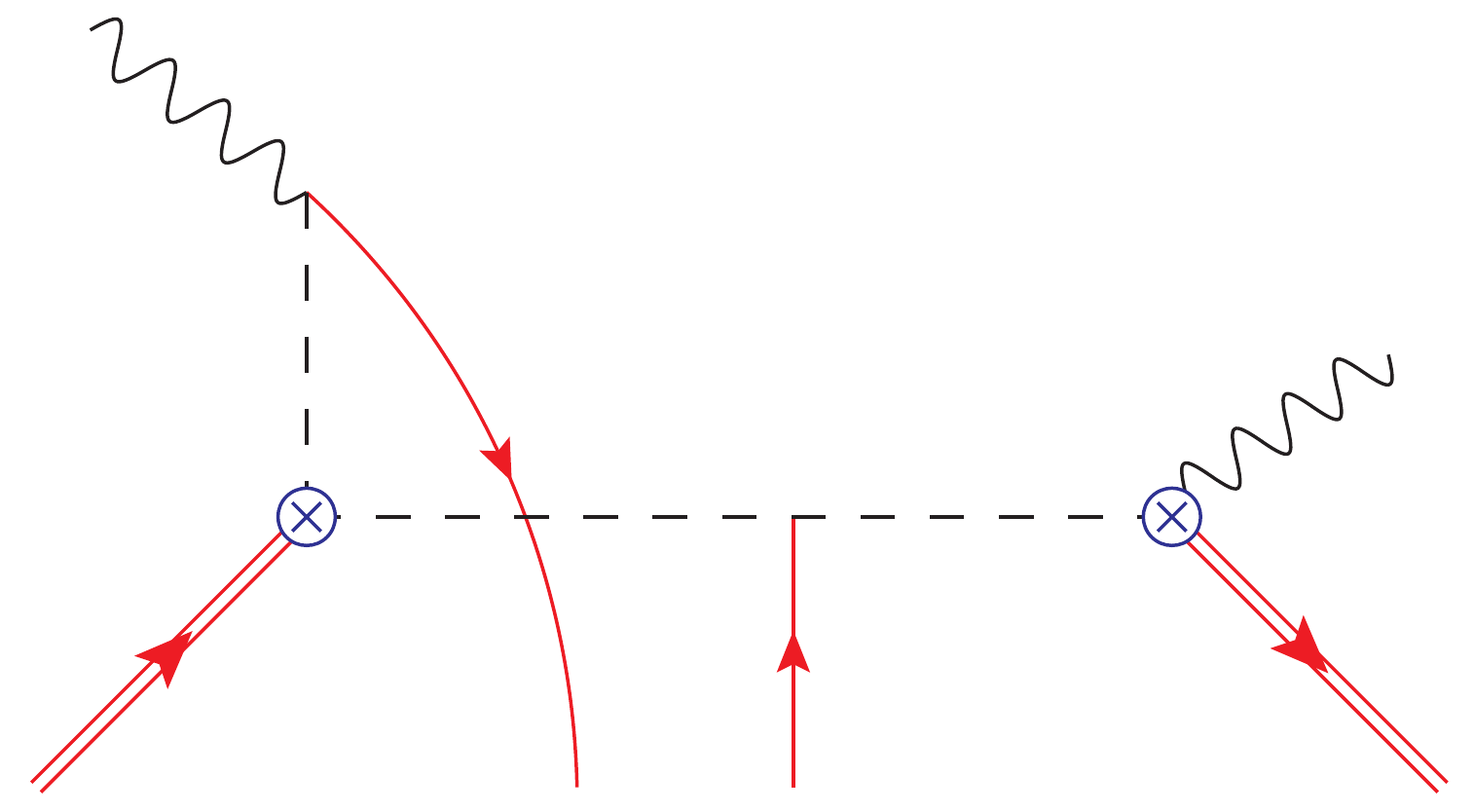}\includegraphics[scale=0.4]{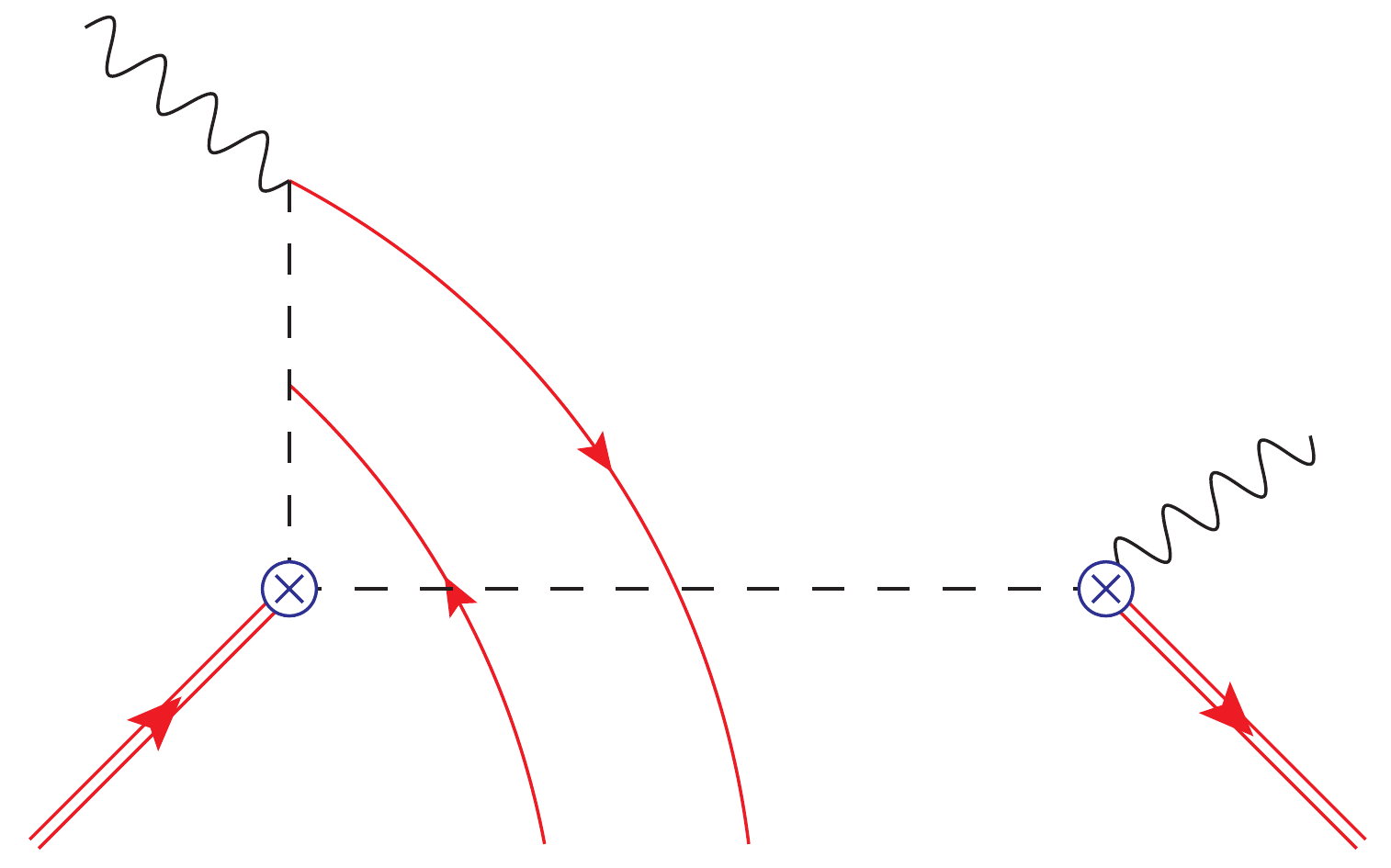}
    \caption{Diagrams arising from the matching of the two ${\cal O}_{7\gamma} - {\cal O}_{8g}$ contributions onto HQET. Red indicates soft fields. Integrating out (anti-) hard-collinear fields leads to non-localities which are denoted by  dashed lines. 
   }  \label{Fig:diagrams0708soft}
\end{center}
\end{figure*}

Within  the ${\cal O}_{7\gamma} - {\cal O}_{8g}$  contribution, there are three cut diagrams. Maintaining the same notation as in Ref.~\cite{Benzke:2010js}, we have for the two cuts with the hard-collinear gluon diagrams (see left diagrams in Figs.~\ref{Fig:diagrams0708} and~\ref{Fig:diagrams0708soft})

\begin{align}
    d\Gamma_{78}^{(b)} &= -\frac{1}{m_b} \,\text{Re} \big[ \hat \Gamma_{7 8} \big] d\Lambda_{\alpha \beta}\, 16\pi\alpha_s\, e_q\, m_b\, n\cdot q\,
(g^{\alpha \beta}_\perp  +  i \epsilon_\perp^{\alpha\beta})\,\text{Re}\! \int d\omega \delta( \omega + m_b - n\cdot q) \nonumber \\
    &\phantom{=\,}\,\times  \int\frac{d\omega_1}{\omega_1 + \bar n \cdot q + i \epsilon}\,  \frac{d \omega_2}{\omega_2 - i \epsilon} \left [\bar g_{78} (\omega, \omega_1,\omega_2,\mu) - \bar g_{78}^\text{cut} (\omega, \omega_1,\omega_2,\mu)\right]\,.
\end{align}
Here the hadronic functions $g_{7 8}$ are defined exactly the same way as already known from the case $B\rightarrow X_s \gamma$.
\begin{align}  
  &\bar g_{78}(\omega,\omega_1,\omega_2,\mu) 
   = \int\frac{dr}{2\pi}\,e^{-i\omega_1 r}\!\int\frac{du}{2\pi}\,e^{i\omega_2 u}\!
    \int\frac{dt}{2\pi}\,e^{-i\omega t} \nonumber \\
   &\times 
    \frac{\langle\bar B| \big(\bar h S_n\big)(tn)\,T^A\,
          \overline{\Gamma}_n\,\big(S_n^\dagger s\big)(un)
          \big(\bar s S_{\bar n}\big)(r\bar n)\,\Gamma_{\bar n}\,
          \big(S_{\bar n}^\dagger S_{n}\big)(0)\,T^A          
          \big(S_n^\dagger h\big)(0) |\bar B\rangle}{2M_B} \nonumber \\
   &\bar g_{78}^{\rm cut}(\omega,\omega_1,\omega_2,\mu) 
    = \int\frac{dr}{2\pi}\,e^{-i\omega_1 r}\!
    \int\frac{du}{2\pi}\,e^{i\omega_2 u}\!
    \int\frac{dt}{2\pi}\,e^{-i\omega t} \nonumber \\
   & \times 
    \frac{\langle\bar B| \big(\bar h S_n\big)(tn)\,T^A\,
          \overline{\Gamma}_n\,\big(S_n^\dagger s\big)((t+u)n)
          \big(\bar s S_{\bar n}\big)(r\bar n)\,\Gamma_{\bar n}\,
          \big(S_{\bar n}^\dagger S_{n}\big)(0)\,T^A          
          \big(S_n^\dagger h\big)(0) |\bar B\rangle}{2M_B} \,,
    \label{eq:g78def1} 
\end{align}
{where $S_n$ and $S_{\bar n}$ are soft Wilson lines connecting the soft fields in the matrix element and thereby ensuring gauge invariance. The exact space-time structure of the operator is depicted in the left of Fig.~\ref{Fig:diagrams0708soft}.}
However, for the cut diagram with an anti-hard-collinear gluon (see right  diagrams in Fig.~\ref{Fig:diagrams0708}
and~\ref{Fig:diagrams0708soft}), we obtain
\begin{align}
    d\Gamma_{78}^{(c)} &= \frac{1}{m_b} \text{Re} \big[ \hat \Gamma_{7 8} \big]d\Lambda_{\alpha \beta}\, 4\pi\alpha_s\, m_b\, n\cdot q\,
(g^{\alpha \beta}_\perp - i \epsilon_\perp^{\alpha \beta}) \,\text{Re}\!\int d\omega \delta( \omega + m_b - n\cdot q)  \nonumber\\
&\phantom{=\,}\,\times  \int\frac{d\omega_1}{\omega_1 - \omega_2 + \bar n \cdot q + i \epsilon}\, d \omega_2   
    \Big [ \left(\frac{1}{\omega_1 +\bar n \cdot q+ i \epsilon} + \frac{1}{\omega_2 - \bar n \cdot q - i \epsilon} \right) g_{78}^{(1)} (\omega, \omega_1,\omega_2,\mu)  \nonumber \\
    &\phantom{\times\Big[ }- \left(\frac{1}{\omega_1 + \bar n\cdot q + i
\epsilon} - \frac{1}{\omega_2 - \bar n \cdot q - i \epsilon} \right)
g_{78}^{(5)} (\omega, \omega_1,\omega_2,\mu)\Big]\,.
\end{align}
Again we find the same shape functions which are defined as
\begin{align} 
   & g_{78}^{(1)}(\omega,\omega_1,\omega_2,\mu) 
   = \int\frac{dr}{2\pi}\,e^{-i\omega_1 r}\!
    \int\frac{du}{2\pi}\,e^{i\omega_2 u}\!
    \int\frac{dt}{2\pi}\,e^{-i\omega t} \nonumber\\
   &\times 
    \frac{\langle\bar B| \big(\bar h S_n\big)(tn) 
          \big(S_n^\dagger S_{\bar n}\big)(0)\,T^A\,
          \slashed{\bar n} (1+\gamma_5)\,
          \big(S_{\bar n}^\dagger h\big)(0)\,{\bf T} 
          \sum{}_q\,e_q\,\big(\bar q S_{\bar n}\big)(r\bar n)\,
          \slashed{\bar n} \,T^A
          \big(S_{\bar n}^\dagger q\big)(u\bar n)
          |\bar B\rangle}{2M_B} \,, \nonumber\\
   &g_{78}^{(5)}(\omega,\omega_1,\omega_2,\mu) 
   = \int\frac{dr}{2\pi}\,e^{-i\omega_1 r}\!
    \int\frac{du}{2\pi}\,e^{i\omega_2 u}\!
    \int\frac{dt}{2\pi}\,e^{-i\omega t} \nonumber \\
   &\times 
    \frac{\langle\bar B| \big(\bar h S_n\big)(tn) 
          \big(S_n^\dagger S_{\bar n}\big)(0)\,T^A
          \slashed{\bar n} (1+\gamma_5)\,
          \big(S_{\bar n}^\dagger h\big)(0)\,{\bf T}
          \sum{}_q\,e_q\,\big(\bar q S_{\bar n}\big)(r\bar n)\,
          \slashed{\bar n}\gamma_5T^A
          \big(S_{\bar n}^\dagger q\big)(u\bar n)
          |\bar B\rangle}{2M_B} \,.\label{eq:g78def2}
\end{align}
{It is clear that the difference to the radiative decay is introduced through the non-vanishing $\bar n\cdot q$ that shifts the small component of the anti-hard-collinear propagator, which corresponds to the anti-hard-collinear jet function. With the same argument, we can already see that the direct contributions will not be affected in such a way, since $\bar n\cdot q$ is suppressed relative to the large component of any hard-collinear propagator.}

\begin{figure*}[t!]
\begin{center}
\includegraphics[scale=0.4]{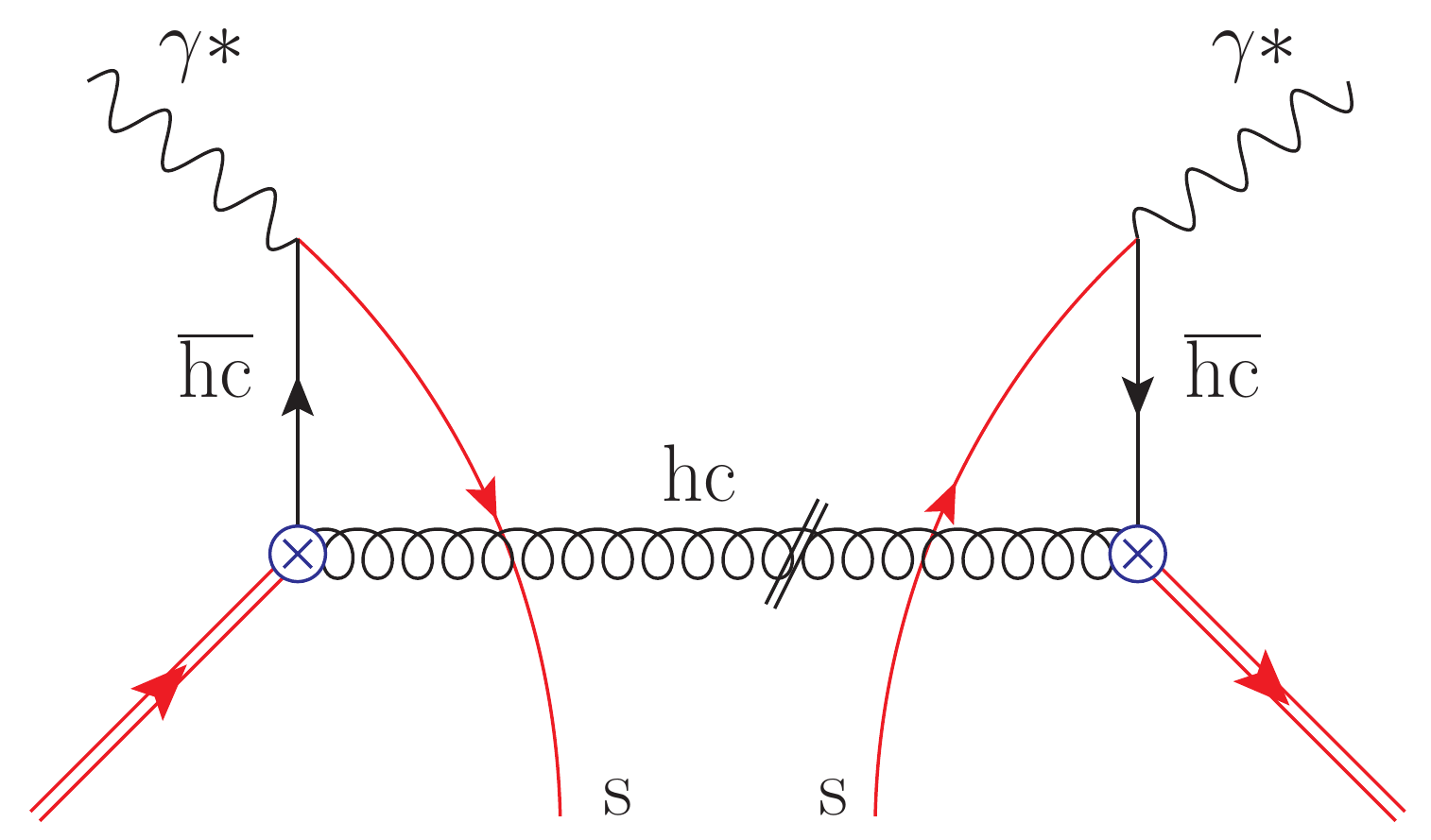}\includegraphics[scale=0.4]{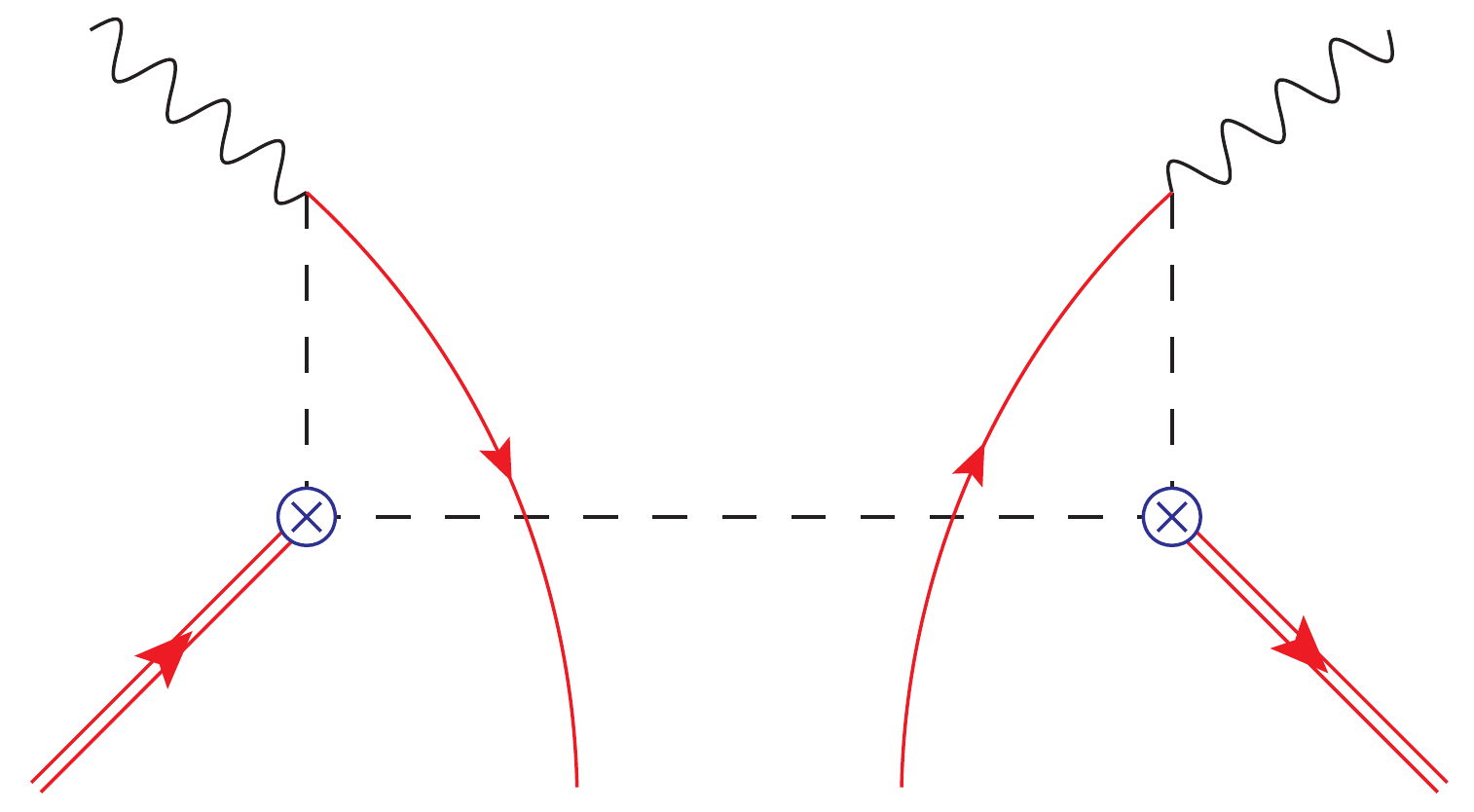}
    \caption{The cut diagram arising from the matching of the  ${\cal O}_{8g} - {\cal O}_{8g}$ contribution onto SCET
    (left) and onto HQET (right).  Red indicates soft fields, black (anti-) hard-collinear fields. Hard fields are already integrated out. Dashed lines correspond to non-localities.} \label{Fig:diagrams0808}
\end{center}
\end{figure*}

For the double resolved   ${\cal O}_{8g} - {\cal O}_{8g}$  contribution  involving twice the QCD dipole operator 
(see diagrams in Fig.~\ref{Fig:diagrams0808}) we find
\begin{align}\label{O8O8}
    d\Gamma_{88} &= \frac{1}{m_b} \,\text{Re}\big[\hat \Gamma_{8 8}\big]\,d\Lambda_{\alpha \beta}\,8\pi\alpha_s\,e_s^2\,m_b^2\,
(g^{\alpha \beta}_\perp + i
\epsilon_\perp^{\alpha \beta})  \,\text{Re}\!\int d\omega \delta( \omega + m_b - n\cdot
q) \nonumber \\
    &\phantom{=\,}\,\times \int  \frac{d\omega_1}{\omega_1 + \bar n \cdot q + i \epsilon}\,  \frac{d \omega_2}{\omega_2 +\bar n \cdot q - i \epsilon} \bar g_{88} (\omega, \omega_1,\omega_2,\mu) \,.
\end{align}
Here the shape function $\bar g_{88}$ is again defined as in the radiative decay
\begin{align}
   & \bar g_{88}(\omega,\omega_1,\omega_2,\mu) 
   = \int\frac{dr}{2\pi}\,e^{-i\omega_1 r}\!\int\frac{du}{2\pi}\,e^{i\omega_2 u}\!
    \int\frac{dt}{2\pi}\,e^{-i\omega t} \label{eq:g88def}\\
   & \times 
    \frac{\langle\bar B| \big(\bar h S_n\big)(tn)\,T^A
          \big(S_n^\dagger S_{\bar n}\big)(tn)\,
          \overline{\Gamma}_{\bar n}
          \big(S_{\bar n}^\dagger s\big)(tn+u\bar n)
          \big(\bar s S_{\bar n}\big)(r\bar n)
          \Gamma_{\bar n}
          \big(S_{\bar n}^\dagger S_{n}\big)(0)\,T^A 
          \big(S_n^\dagger h\big)(0)
          |\bar B\rangle}{2M_B} \,. \nonumber
\end{align}
As mentioned already in Section 2.2, there is a subtlety concerning the convolution integral in Eq.~\ref{O8O8}. When calculating the asymptotic behaviour of the soft function for $\omega_{1,2}  \gg \Lambda_{\rm QCD}$ one finds that the convolution integrals are UV divergent \cite{Benzke:2010js}. This divergence is mirrored by an IR divergence in the direct contribution to ${\cal O}_{8g}-{\cal O}_{8g}$. In order to properly define all quantities it is necessary to split the convolution integrals in Eq.~\ref{O8O8} into an UV part with $\omega_{1,2} > \Lambda_{{\rm UV}}$ and an IR part with $\omega_{1,2} < \Lambda_{{\rm UV}}$. In the sum of direct and resolved contributions the divergence cancels, there remains, however, a logarithmic dependence on the parameter $\Lambda_{{\rm UV}}$ {in the perturbative part.}

\begin{figure*}[t!]
\begin{center}
\includegraphics[scale=0.4]{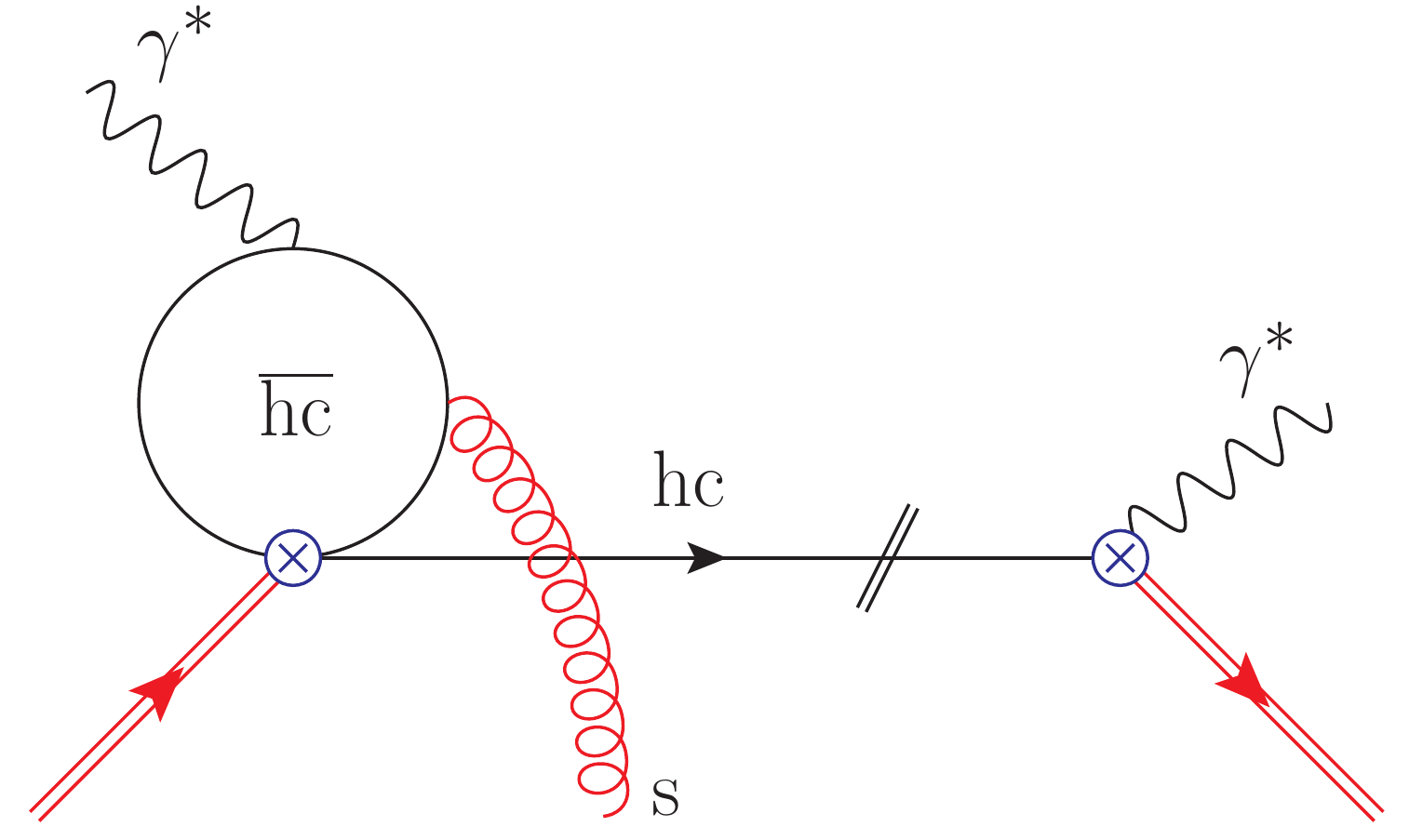}\includegraphics[scale=0.4]{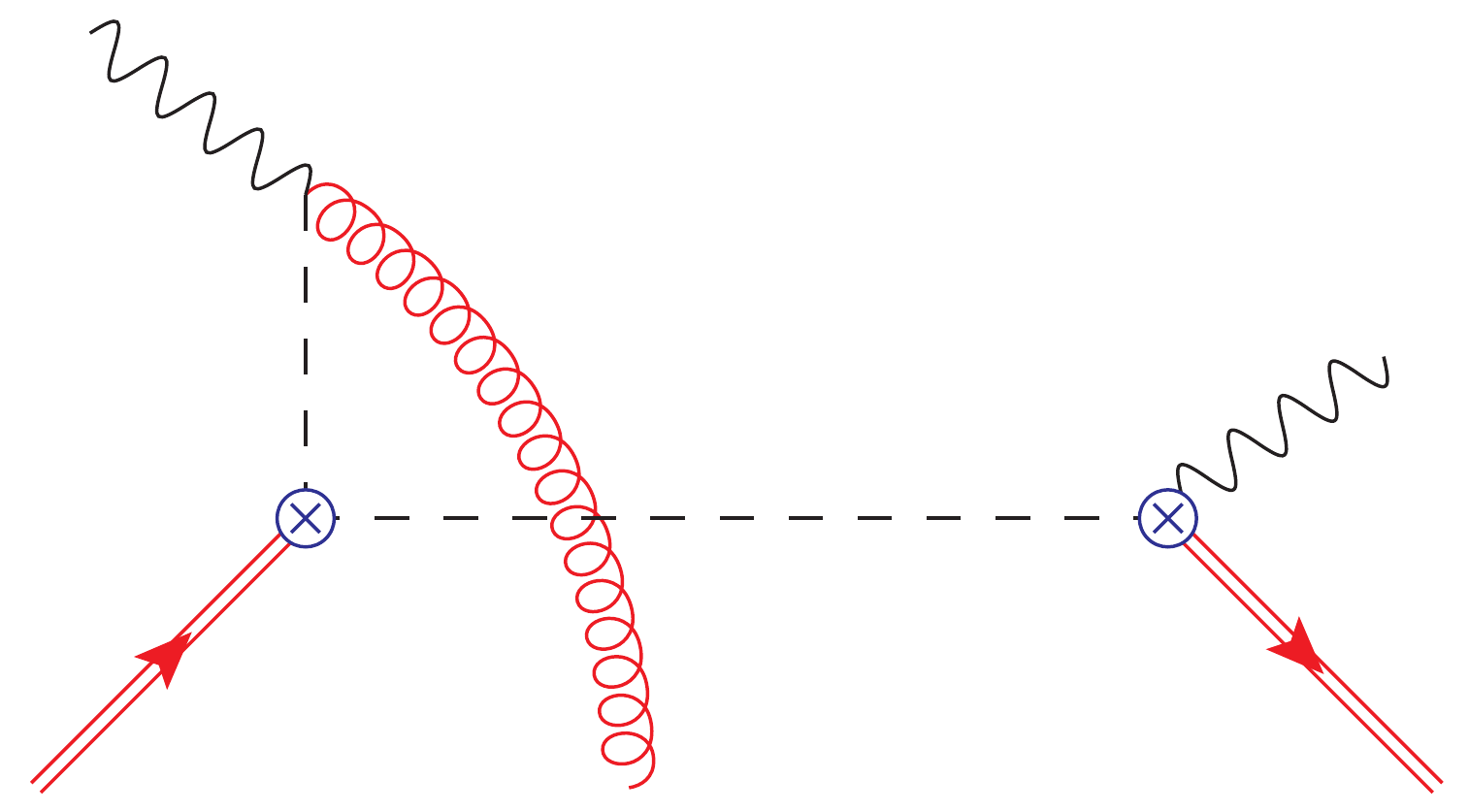}
    \caption{The cut diagram arising from the matching of the  ${\cal O}_{1} - {\cal O}_{7\gamma}$ contribution onto SCET
    (left) and onto HQET (right).  Red indicates soft fields, black (anti-) hard-collinear fields. Hard fields are already integrated out. Dashed lines correspond to non-localities.}  \label{Fig:diagrams0107}
\end{center}
\end{figure*}

For the ${\cal O}_1 - {\cal O}_{7\gamma}$ contribution (see Fig.~\ref{Fig:diagrams0107})  we have explicitly derived
\begin{align}
d\Gamma_{1 7} =& \frac{1}{m_b}\,\text{Re}\Big[\hat\Gamma_{1 7}\frac{-\lambda_t^*\lambda_c}{|\lambda_t|^2}\Big]\,
d\Lambda_{\alpha\beta}\, e_c\, (n\cdot q)^2\,
\text{Re}\int d\omega \delta( \omega + m_b - n\cdot q)\int d\omega_1 \frac{1}{\omega_1 + i \epsilon}\nonumber\\
&\times\frac{1}{\omega_1}\left[
(\nb\cdot q+\omega_1)\left(1-F\left(\frac{m_c^2}{n\cdot q(\nb\cdot q+\omega_1)}\right)\right)
-\nb\cdot q \left(1-F\left(\frac{m_c^2}{n\cdot q\nb\cdot q}\right)\right) \right.\nn\\
&\left.-\nb\cdot q \left( G\left(\frac{m_c^2}{n\cdot q(\nb\cdot q+\omega_1)}\right) - G\left(\frac{m_c^2}{n\cdot q\nb\cdot q}\right)\right)\right]\nonumber\\
&\times\int \!\frac{dt}{2\pi} e^{-i \omega t}\!\!\int \!\frac{dr}{2\pi} e^{-i \omega_1 r}
\frac{\langle B | \bar h(nt) \slashed{\bar n} [1+\gamma^5]\frac{i}{2} [\gamma^\mu_\perp, \gamma^\beta_\perp]\gamma_\perp^\alpha \bar{n}^\kappa g G_{\mu \kappa}(\bar n r)  h(0) | B \rangle}{2M_B}\,.
\end{align}
The decomposition of the Lorentz structure has been done above (see  Section~\ref{sec:example}).  
\begin{align}
d\Gamma_{1 7} =& \frac{1}{m_b}\,\text{Re}\Big[\hat\Gamma_{1 7}\frac{-\lambda_t^*\lambda_c}{|\lambda_t|^2}\Big]
\frac{\alpha}{24\pi^3}\,dn\cdot qd\bar n\cdot q\,\frac{(n\cdot q)^3}{\bar n\cdot q}\,
\text{Re}\int d\omega \delta( \omega + m_b - n\cdot q)\int d\omega_1 \frac{1}{\omega_1 + i \epsilon}\nonumber\\
&\times\frac{1}{\omega_1}\left[
(\nb\cdot q+\omega_1)\left(1-F\left(\frac{m_c^2}{n\cdot q(\nb\cdot q+\omega_1)}\right)\right)
-\nb\cdot q \left(1-F\left(\frac{m_c^2}{n\cdot q\nb\cdot q}\right)\right) \right.\nn\\
&\left.-\nb\cdot q \left( G\left(\frac{m_c^2}{n\cdot q(\nb\cdot q+\omega_1)}\right) - G\left(\frac{m_c^2}{n\cdot q\nb\cdot q}\right)\right)\right]
\,g_{17}(\omega,\omega_1,\mu)\,,
\end{align}
with
\begin{eqnarray}
   g_{17}(\omega,\omega_1,\mu) 
   &=& \int\frac{dr}{2\pi}\,e^{-i\omega_1 r}\!
    \int\frac{dt}{2\pi}\,e^{-i\omega t} \\
   &&\times \frac{\langle\bar B| \big(\bar h S_n\big)(tn)\,
    \slashed{\bar n} (1+\gamma_5) \big(S_n^\dagger S_{\bar n}\big)(0)\,
    i\gamma_\alpha^\perp\bar n_\beta\,
    \big(S_{\bar n}^\dagger\,g G_s^{\alpha\beta} S_{\bar n} 
    \big)(r\bar n)\,
    \big(S_{\bar n}^\dagger h\big)(0) |\bar B\rangle}{2M_B} \,.
    \nonumber
\end{eqnarray}

Finally some remarks are in order: 

\begin{itemize}
\item   Having listed  our results for the triple differential decay rate above with the calculated phase space inserted, we find that there is no odd term in the  variable $z$. Thus,  there is no resolved contribution to the forward-backward asymmetry in the first subleading order. 
\item  Strictly speaking the CP averaging with the real part prescription is only valid because no linear term in $z$ appears, as for the CP conjugated rate we would have to replace $z\rightarrow -z$.
\item  All diagrams show that if we considered  the lepton momenta as hard, the resolved contributions would not exist. The hard momentum of the leptons would imply  also a hard momentum of the intermediate parton. The latter would be integrated out at the hard scale and the virtual photon would be directly connected to the effective electroweak interaction vertex. 
\item  As the various results show, the shape function is non-local in both light cone directions. Thus, the resolved contributions stay non-local even when the hadronic mass cut is relaxed. 
{In that case $n\cdot P_X=M_B-n\cdot q$ is not necessarily small anymore. We can therefore expand the shape function in powers of $\Lambda_\text{QCD}/(m_b-n\cdot q)$ which leads to a series of matrix elements that are local on the $n$-direction. However, the non-locality in the $\bar n$ direction is retained.}
In this sense the resolved contributions represent an irreducible uncertainty within the  inclusive decay $\bar B \to X_s \ell^+\ell^-$. 
\end{itemize}

\section{Numerical Analysis } \label{sec:numerics}

{First we discuss our input parameters. For the bottom-quark mass we use the low-scale subtracted heavy quark mass defined in the shape-function scheme: $m_b = 4.65\, \text{GeV}$~\cite{ Bosch:2004th}. However, we vary the mass between the running mass in the MS scheme, $\overline{m}_b^{\rm MS}(m_b) = 4.2\, \text{GeV}$, and the pole mass, $m_b^{\rm pole} = 4.8\, \text{GeV}$. The charm-quark mass enters as a running mass in the charm-penguin diagrams with a soft-gluon emission (within the interference of ${\cal O}_1$ with ${\cal O}_{7\gamma}$). 
This diagram lives at the hard-collinear scale $\mu_{\rm hc}  = \sqrt{m_b \Lambda_{\rm QCD}}$, thus, we choose $\overline{m}_c^{\rm MS}(\mu_{\rm hc} = 1.5\, \text{GeV}) = 1.131\, \text{GeV}$. The variation of $m_b$ implies 
$1.45\, \text{GeV} < \mu_{\rm hc}  <  1.55\, \text{GeV} $.  We conservatively  vary the hard-collinear scale
between $\mu_{\rm hc} = 1.4\, \text{GeV}$ and $\mu_{\rm hc} = 1.6\, \text{GeV}$.

Regarding the HQET parameters we adopt  the choices of Ref.~\cite{Benzke:2010js}: We use $\lambda_2=(0.12\pm 0.02)$\,GeV$^2$ . For the first inverse moment of the $B$-meson light-cone distribution amplitude, we take the range 
$0.25\,\mbox{GeV}<\lambda_B<0.75\,\mbox{GeV}$. For the parameter $F$ we use the relation $F=f_B\sqrt{M_B}$, and with  $f_B=(193\pm 10)$\,MeV we finally obtain $0.177\,{\rm GeV}^3<F^2<0.217\,{\rm GeV}^3$.

We use NLO Wilson coefficients. However, in the BBL basis used in our analysis, the coefficients $C_{7\gamma}$ and $C_{8}$ are only known to LO. We crosschecked the numerical impact compared to using the 
CMM basis~\cite{Chetyrkin:1996vx}  for which all coefficients are known at least to NLO accuracy. We find that the numerical effect of the change is negligible in view of the other uncertainties within our analysis.}

\subsection{\texorpdfstring{Interference of ${\cal O}_1$ with ${\cal O}_{7\gamma}$}{Interference of Q1 with Q7}}

We are interested in the relative magnitude of the resolved contributions compared to the total decay rate, {i.e. the leading direct contributions to the decay rate which one also would consider when the decay rate was calculated within the OPE}
\begin{equation}
{\mathcal F}(q_\mathrm{min}^2,q_\mathrm{max}^2,M_{X,\mathrm{max}}^2)
=\frac{\Gamma_\mathrm{resolved}(q_\mathrm{min}^2,q_\mathrm{max}^2,M_{X,\mathrm{max}}^2)}{\Gamma_\mathrm{OPE}(q_\mathrm{min}^2,q_\mathrm{max}^2,M_{X,\mathrm{max}}^2)}\,,
\label{eqn:def}
\end{equation}
where the rate $\Gamma_\mathrm{OPE}$ is given by
\begin{align}
\Gamma_\mathrm{OPE} =&\, \frac{G_F^2\alpha m_b^5}{32\pi^4}\,|V_{tb}^*V_{ts}|^2\frac{1}{3}\frac{\alpha}{\pi}\int\frac{d\bar n\cdot q}{\bar n \cdot q}
\left(1-\frac{\bar n\cdot q}{m_b}\right)^2\nonumber\\
&\,\Bigg[ C_{7\gamma}^2\Bigg(1+\frac{1}{2}\frac{\bar n\cdot q}{m_b}\Bigg)
+(C_9^2+C_{10}^2)\Bigg(\frac{1}{8}\frac{\bar n\cdot q}{m_b}+\frac{1}{4}\left(\frac{\bar n\cdot q}{m_b}\right)^2\Bigg)
+C_{7\gamma}C_9\frac{3}{2}\frac{\bar n\cdot q}{m_b}\Bigg]\nonumber\\
\equiv&\,\frac{G_F^2\alpha m_b^5}{32\pi^4}\,|V_{tb}^*V_{ts}|^2\frac{1}{3}\frac{\alpha}{\pi}\, C_{\rm OPE}\,.
\end{align}
The last line defines the quantity $C_{\rm OPE}$. The integration limits are specified below. 

The first term in the square brackets is the leading power in the $1/m_b$ expansion and corresponds to the direct contribution due to the interference of ${\cal O}_{7\gamma}$ with itself. The other terms are formally suppressed { in the shape function region in  which we evaluate these direct contributions}. But 
  the large magnitude of the Wilson coefficients $|C_{9/10}|\sim 13|C_{7\gamma}|$ necessitates their inclusion into our uncertainty.

For the resolved contribution from the  interference of ${\cal O}_1$ with ${\cal O}_{7\gamma}$ we find 
\begin{align}
&{\mathcal F}_{17}
=\frac{1}{m_b^4}\frac{C_1(\mu)C_{7\gamma}(\mu)}{C_{\rm OPE}} e_c\,
\mathrm{Re} \int d\nb\cdot q\,dn\cdot q\, \frac{(n\cdot q)^3}{\nb\cdot q}
\int_{-p_+}^{\bar{\Lambda}} d\omega\,\delta(m_b - n\cdot q +\omega)
\int_{-\infty}^{+\infty}\frac{d\omega_1}{\omega_1+\ie}\nn\\
&\frac{1}{\omega_1}\left[
(\nb\cdot q+\omega_1)\left(1-F\left(\frac{m_c^2}{n\cdot q(\nb\cdot q+\omega_1)}\right)\right) 
-\nb\cdot q \left(1-F\left(\frac{m_c^2}{n\cdot q\nb\cdot q}\right)\right)\right.\nn\\
&\left.-\nb\cdot q \left( G\left(\frac{m_c^2}{n\cdot q(\nb\cdot q+\omega_1)}\right) - G\left(\frac{m_c^2}{n\cdot q\nb\cdot q}\right)\right)\right]
g_{17}(\omega,\omega_1,\mu)\,,
\label{eqn:f17}
\end{align}
where we have neglected terms proportional to $V_{ub}$. Here   $p_+ \equiv n\cdot p = m_b - n\cdot q$, $\bar \Lambda = M_B - m_b$, and 
the penguin functions $F$ and $G$ are defined in equation \ref{FF}.

The integration limits of the $n\cdot q$ and $\bar n\cdot q$ can be read of Fig.~\ref{Fig:scaling_law}. To order 
$\lambda^2$ they are
\begin{align}
&\int_{\frac{q_\mathrm{min}^2}{M_B}}^{\frac{q_\mathrm{min}^2}{M_B}+\frac{M_{X,\mathrm{max}}^2q_\mathrm{min}^2}{M_B^3}}d\nb\cdot q
\int_{\frac{q_\mathrm{min}^2}{\nb\cdot q}}^{M_B} dn\cdot q \nn\\
+&\int_{\frac{q_\mathrm{min}^2}{M_B}+\frac{M_{X,\mathrm{max}}^2q_\mathrm{min}^2}{M_B^3}}^{\frac{q_\mathrm{max}^2}{M_B}}d\nb\cdot q
\int_{M_B-\frac{M_{X,\mathrm{max}}^2}{M_B-\nb\cdot q}}^{M_B} dn\cdot q \nn \\
+&\int_{\frac{q_\mathrm{max}^2}{M_B}}^{\frac{q_\mathrm{max}^2}{M_B}+\frac{M_{X,\mathrm{max}}^2q_\mathrm{max}^2}{M_B^3}}d\nb\cdot q
\int_{M_B-\frac{M_{X,\mathrm{max}}^2}{M_B-\nb\cdot q}}^{\frac{q_\mathrm{max}^2}{\nb\cdot q}}dn\cdot q
\end{align}
Since the integrand of the $\nb\cdot q$ integration is not singular, the first and third line do not give a leading power contribution.
Note, that the integration limits of $n\cdot q$ are of $\mathcal{O}(1)$ in all terms, as they have to be, but the integration region is only of $\mathcal{O}(\lambda)$. To illustrate  this we can substitute $n\cdot q\rightarrow m_b-p_+$. For convenience, we reverse the sign $p_+\rightarrow -p_+$. The integration in the first line of equation (\ref{eqn:f17}) can then be written as
\begin{equation}
\dots\int_{\frac{q_\mathrm{min}^2}{M_B}}^{\frac{q_\mathrm{max}^2}{M_B}}d\nb\cdot q 
\int_{\bar{\Lambda}-\frac{M_{X,\mathrm{max}}^2}{M_B-\nb\cdot q}}^{\bar{\Lambda}} dp_+
\, \frac{(m_b+p_+)^3}{\nb\cdot q} \int_{p_+}^{\bar{\Lambda}} d\omega\,\delta(\omega-p_+)\dots\,.
\end{equation}
Changing the order of the $\omega$ and $p_+$ integrations,
\begin{equation}
\int_{\bar{\Lambda}-\frac{M_{X,\mathrm{max}}^2}{M_B-\nb\cdot q}}^{\bar{\Lambda}} dp_+
\int_{p_+}^{\bar{\Lambda}} d\omega
=\int_{\bar{\Lambda}-\frac{M_{X,\mathrm{max}}^2}{M_B-\nb\cdot q}}^{\bar{\Lambda}} d\omega
\int_{\bar{\Lambda}-\frac{M_{X,\mathrm{max}}^2}{M_B-\nb\cdot q}}^{\omega} dp_+
\end{equation}
and performing  the $p_+$ integration to eliminate the $\delta$ distribution, yields for ${\mathcal F}_{17}$ 
\begin{align}
&{\mathcal F}_{17}
=\frac{1}{m_b^4}\frac{C_1(\mu)C_{7\gamma}(\mu)}{C_{\rm OPE}} e_c\,
\mathrm{Re} \int_{\frac{q_\mathrm{min}^2}{M_B}}^{\frac{q_\mathrm{max}^2}{M_B}}d\nb\cdot q
\, \frac{m_b^3}{\nb\cdot q}
\int_{-\infty}^{+\infty}\frac{d\omega_1}{\omega_1+\ie}\,\nn\\
&\frac{1}{\omega_1}\left[
(\nb\cdot q+\omega_1)\left(1-F\left(\frac{m_c^2}{m_b(\nb\cdot q+\omega_1)}\right)\right) 
-\nb\cdot q \left(1-F\left(\frac{m_c^2}{m_b\nb\cdot q}\right)\right)\right.\nn\\
&\left.-\nb\cdot q \left( G\left(\frac{m_c^2}{m_b(\nb\cdot q+\omega_1)}\right) - G\left(\frac{m_c^2}{m_b\nb\cdot q}\right)\right)\right]
\int_{\bar{\Lambda}-\frac{M_{X,\mathrm{max}}^2}{M_B-\nb\cdot q}}^{\bar{\Lambda}} d\omega\, g_{17}(\omega,\omega_1,\mu)\,.
\label{eqn:f17mb}
\end{align}
From this we define
\begin{equation}
h_{17}(M_{X,\mathrm{max}},\omega_1,\mu) = \int_{\bar{\Lambda}-\frac{M_{X,\mathrm{max}}^2}{M_B-\nb\cdot q}}^{\bar{\Lambda}} d\omega\, g_{17}(\omega,\omega_1,\mu)\,.
\end{equation}
Since the soft function only has support for $\omega\sim\Lambda_\mathrm{QCD}$ we can take the limit  $M_{X,\mathrm{max}}\rightarrow M_B$ to get\footnote{Note, that this does not work for $g_{88}$, since we would put two light quark fields at a light-like distance, which yields a divergent propagator.}
\begin{equation}
h_{17}(\omega_1,\mu) = \int\frac{dr}{2\pi}\,e^{-i\omega_1r}\frac{\braB \bar{h}(0)\nbs i \gamma_\alpha^\perp\nb_\beta gG^{\alpha\beta}(r\nb)h(0)\ketB}{2M_B}
\end{equation}
{Knowing the explicit form of the HQET matrix element we can derive general properties of the shape function $h_{17}$. Following the arguments given in Ref.~\cite{Benzke:2010js},
one can derive from  PT invariance that the function is real and even in $\omega_1$. 
One can also explicitly derive the general normalization of the soft function
\begin{equation}
\int^\infty_{-\infty} d\omega_1 h_{17}(\omega_1,\mu) = 2\, \lambda_2\,. 
\end{equation}
Finally, the  soft function $h_{17}$ should not have any significant structure (maxima or zeros) outside the hadronic range, and  the values of $h_{17}$ should be within the hadronic range.}  

\begin{figure}[hpt]
\centering
\includegraphics[width=120mm]{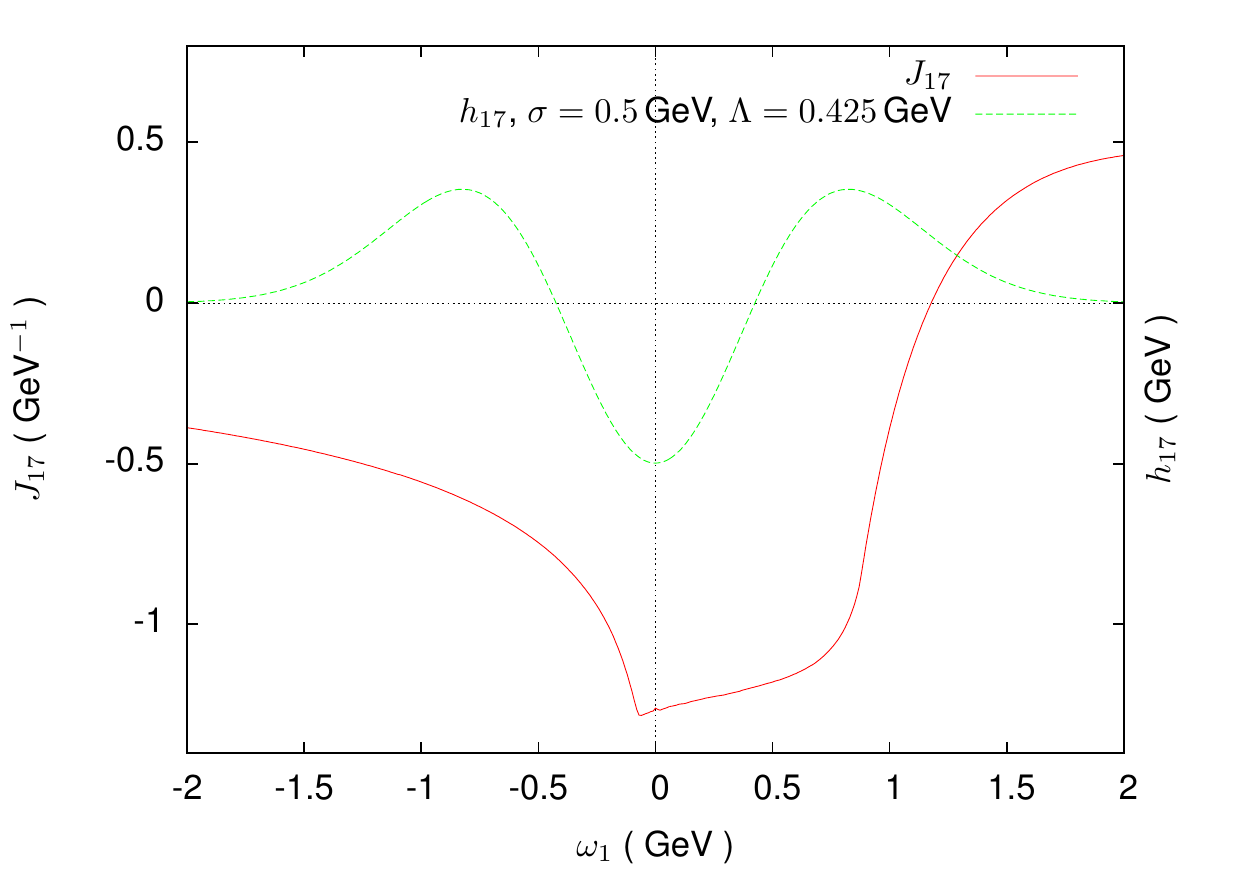}
\caption{$J_{17}$ for $q_\mathrm{min}^2=1\,$GeV$^2$ and $q_\mathrm{max}^2=6\,$GeV$^2$, together with the model function of equation \ref{eqn:h17b}.}\label{fig:J17}
\end{figure}

In summary, we can write the relative contribution due to the interference of ${\cal O}_1$ with ${\cal O}_{7\gamma}$ as
\begin{align}
&{\mathcal F}_{17}
=\frac{1}{m_b}\frac{C_1(\mu)C_{7\gamma}(\mu)}{C_{\rm OPE}} e_c\,
\int_{-\infty}^{+\infty}d\omega_1\,
J_{17}(q_\mathrm{min}^2,q_\mathrm{max}^2,\omega_1)\,
h_{17}(\omega_1,\mu)
\label{eqn:full}
\end{align}
with
\begin{align}
&J_{17}(q_\mathrm{min}^2,q_\mathrm{max}^2,\omega_1) = 
\mathrm{Re} \frac{1}{\omega_1+\ie}
\int_{\frac{q_\mathrm{min}^2}{M_B}}^{\frac{q_\mathrm{max}^2}{M_B}} \frac{d\nb\cdot q}{\nb\cdot q}\,
\frac{1}{\omega_1} \nn\\
&\left[
(\nb\cdot q+\omega_1)\left(1-F\left(\frac{m_c^2}{m_b(\nb\cdot q+\omega_1)}\right)\right) 
-\nb\cdot q \left(1-F\left(\frac{m_c^2}{m_b\nb\cdot q}\right)\right)\right.\nn\\
&\left.-\nb\cdot q \left( G\left(\frac{m_c^2}{m_b(\nb\cdot q+\omega_1)}\right) - G\left(\frac{m_c^2}{m_b\nb\cdot q}\right)\right)\right]\,.
\end{align}
For the standard value of $q_\mathrm{min}^2$ and $q_\mathrm{max}^2$ the function $J_{17}$ is plotted in 
Fig.~\ref{fig:J17}.
It is largest around $\omega_1=0$.

As a first trial for a model function for $h_{17}$, we use a Gaussian.
\begin{equation}
h_{17}(\omega_1)=\frac{2\lambda_2}{\sqrt{2\pi}\sigma}e^{-\frac{\omega_1^2}{2\sigma^2}}\,,
\end{equation}
with $\sigma =0.5\,$GeV as typical hadronic scale. This model function has all properties one  derives from the explicit 
HQET matrix element. Calculating the convolution integral, we find 
\begin{align}
{\mathcal F}_{17\mathrm{Gaussian}}=\frac{1}{m_b}\frac{C_1(\mu)C_{7\gamma}(\mu)}{C_{\rm OPE}} e_c
\,(-0.252
\,\mathrm{GeV})\,.
\end{align}
Using a smaller $\sigma=0.1\,$GeV leads to $-0.304\,$GeV. We can express our numbers in percentages
\begin{align}
{\mathcal F}_{17\mathrm{exp}}\approx&\, +1.9\,\%\nonumber\\
\end{align}
Using a Gaussian for the soft function only yields negative numbers (positive percentages) for the expression in the square brackets.
Thus, this model function does not lead to a conservative bound on the size of ${\mathcal F}_{17}$.

Using the same function as in Ref.~\cite{Benzke:2010js} 
\begin{equation}
h_{17}(\omega_1)=\frac{2\lambda_2}{\sqrt{2\pi}\sigma}\frac{\omega_1^2-\Lambda^2}{\sigma^2-\Lambda^2}e^{-\frac{\omega_1^2}{2\sigma^2}}\,,
\label{eqn:h17b}
\end{equation}
we  also get positive numbers for this expression. If $\Lambda$ and $\sigma$ are chosen of order $\Lambda_{\text QCD}$ again all general properties derived for $h_{17}$ are fulfilled.   For a parameter choice of $\sigma=0.5\,$GeV and $\Lambda=0.425\,$GeV one finds
\begin{equation} \label{convolution17}
{\mathcal F}_{17}=\frac{1}{m_b}\frac{C_1(\mu)C_{7\gamma}(\mu)}{C_{\rm OPE}} e_c\,(+0.075\,\mathrm{GeV}).
\end{equation}
For a different parameter choice, $\Lambda=0.575\,$GeV, on the other hand
\begin{equation}
{\mathcal F}_{17}=\frac{1}{m_b}\frac{C_1(\mu)C_{7\gamma}(\mu)}{C_{\rm OPE}} e_c\,(-0.532\,\mathrm{GeV}).
\end{equation}
which leads us to conservative estimate
\begin{equation}
{\mathcal F}_{17}\in [-0.5,+3.4]\,\%\,.
\end{equation}
By reducing the separation between $\Lambda$ and $\sigma$ one could reach larger values, but it would also increase the values of the soft function to outside the hadronic range.

As mentioned in the introduction, for the decay $\bar B \to X_s \gamma$,  it is possible
to expand this non-local  contribution to  local operators  {\it if} the charm quark is treated as heavy.
The first term in this expansion is the dominating one~\cite{Voloshin:1996gw,Ligeti:1997tc,Grant:1997ec,Buchalla:1997ky} which corresponds to the so-called Voloshin term.  
This non-perturbative correction is suppressed by $\lambda_2/m_c^2$.  
But if the charm mass is assumed to scale as 
$m_c^2\sim\Lambda_{\text{QCD}} m_b$, what  seems a more reasonable assumption, the charm penguin contribution must be described by the matrix element of a non-local
operator~\cite{Benzke:2010js}. 

The same can be shown in the decay  $\bar B \to X_s \ell^+\ell^-$.
In Ref.~\cite{Buchalla:1997ky}, the local Voloshin term was  
derived from a local expansion assuming 
{$\Lambda_\text{QCD}m_b/m_c^2$ to be small.}
We can rederive the leading term (according to our power counting) of their result  from our general result above under the following  assumptions.  

Using  a Gaussian as shape function and assuming  this function being narrow enough, 
one can expand  the part of the integrand in square brackets in Eq.~(\ref{eqn:f17mb}) around  $\omega_1=0$\footnote{{ The variable  $(m_b \omega_1) / m_c^2$ corresponds to the parameter $t = k\cdot q / m_c^2$ in Ref.~\cite{Buchalla:1997ky} which is used there as expansion parameter. Note that we have already expanded  in $\bar n\cdot q/m_b$  within the non-local contribution in order to single out the $1/m_b$ term.}}
\begin{align}
\big[\dots\big] =\,& \omega_1^2\bar n\cdot q\,\Bigg[
\frac{1}{2\bar n\cdot q^2} \nonumber\\
-\,&\frac{2m_c^2}{\bar n\cdot q^2}\frac{1}{4m_c^2-m_b\bar n\cdot q}\sqrt{\frac{4m_c^2-m_b\bar n\cdot q}{m_b\bar n\cdot q}}
\arctan\frac{1}{\sqrt{\frac{4m_c^2-m_b\bar n\cdot q}{m_b\bar n\cdot q}}}\Bigg] \nonumber\\
=&\,-\frac{m_b\omega_1^2}{12m_c^2} F_\mathrm{V}(r)\,,
\end{align}
where $ F_\mathrm{V}(r)$ is defined in Eq. (4) of \cite{Buchalla:1997ky} with $r=q^2/(4m_c^2)$ (which is different from the function $F$ defined in  Eq.~\ref{FF}.). This corresponds exactly to the leading power in $1/m_b$ of the Voloshin term for  $\bar B \to X_s \ell^+ \ell^-$ given in Ref.~\cite{Buchalla:1997ky}.  For $F_\mathrm{V}(0) =1$, this results in the Voloshin term for $\bar B \to X_s \gamma$.

Numerically this approach is not advisable.  Evaluating the leading $1/m_b$ Voloshin term yields
\begin{align}
{\mathcal F}_\mathrm{Voloshin,m_b^{-1}}=&\,\frac{1}{m_b}\frac{C_1(\mu)C_{7\gamma}(\mu)}{C_{\rm OPE}} e_c
\int_{\frac{q_\mathrm{min}^2}{M_B}}^{\frac{q_\mathrm{max}^2}{M_B}} \frac{d\nb\cdot q}{\nb\cdot q}\,
\left( -\frac{m_b2\lambda_2}{12m_c^2} \right)
F_\mathrm{V}\left(\frac{m_b\bar n\cdot q}{4m_c^2}\right)\nonumber\\
=&\,\frac{1}{m_b}\frac{C_1(\mu)C_{7\gamma}(\mu)}{C_{\rm OPE}} e_c
\,(-0.306\,\mathrm{GeV})\,.
\end{align}
Compared to our final estimate, we find that the Voloshin term  significantly underestimates the possible charm contributions.

For comparison we finally consider the higher orders in $1/m_b$ of the Voloshin term
derived in Ref.~\cite{Buchalla:1997ky}.  They are given by
\begin{align}
{\mathcal F}_\mathrm{Voloshin}=&\,\frac{1}{m_b}\frac{C_1(\mu)}{C_{\rm OPE}}  e_c
\int_{\frac{q_\mathrm{min}^2}{M_B}}^{\frac{q_\mathrm{max}^2}{M_B}} \frac{d\nb\cdot q}{\nb\cdot q}\,
\left( -\frac{m_b2\lambda_2}{12m_c^2} \right)
F_\mathrm{B.I.}\left(\frac{m_b\bar n\cdot q}{4m_c^2}\right)\nonumber\\
&\,\Bigg[C_{7\gamma}(\mu)\Bigg(1+6\frac{\bar n\cdot q}{m_b}-\left(\frac{\bar n\cdot q}{m_b}\right)^2\Bigg)
+C_9(\mu)\Bigg(2\frac{\bar n\cdot q}{m_b}+\left(\frac{\bar n\cdot q}{m_b}\right)^2\Bigg)\Bigg]\nonumber\\
=&\,\frac{1}{m_b}\frac{C_1(\mu)C_{7\gamma}(\mu)}{C_{\rm OPE}} e_c
\,(+0.481\,\mathrm{GeV})\,.
\end{align}
We note that the higher order in $\bar n\cdot q $ are numerically small but the first subleading $C_9$ is numerically significant  taking into account  $ |C_{9/10}| \sim 13  |C_{7\gamma}|$. We also find that these 
subleading contributions change the sign 
\begin{align}
{\mathcal F}_\mathrm{Voloshin,m_b^{-1}}\approx&\, +1.9\,\%\nonumber\\
{\mathcal F}_\mathrm{Voloshin}\approx&\, -3.0\,\%
\end{align}
Clearly, within the Voloshin term   there is a cancellation between the  $C_{7\gamma}$ and the subleading 
$C_9$  contribution but in our  analysis in which we use $m_c^2 \sim m_b \Lambda_{\rm QCD}$ both  terms get smeared out by different shape functions and, thus, the corresponding uncertainties have to be added up. These  findings call for a calculation of the resolved contributions to order $1/m_b^2$  to collect all numerically relevant contributions~\cite{workinprogress}.

\subsection{\texorpdfstring{Interference of ${\cal O}_{7\gamma}$ with ${\cal O}_{8g}$}{Interference of Q7 with Q8}}

The relative uncertainty due to the interference of ${\cal O}_{7\gamma}$ and ${\cal O}_{8g}$ consists of two contributions ${\mathcal F}_{78}^{(b)}$ and ${\mathcal F}_{78}^{(c)}$. 

{From the explicit form of the shape functions given in Eqs.~\ref{eq:g78def1} and~\ref{eq:g78def2}  it can be deduced  (see Ref.~\cite{Benzke:2010js}) that 
the soft functions $\bar g_{78}$ and $g_{78}^{(1,5)}$ have support for $-\infty<\omega\le\bar\Lambda$ and $-\infty<\omega_{1,2}<\infty$, and 
\begin{equation}\label{g78sym}
   \int_{-\infty}^{\bar\Lambda}\!d\omega 
    \left[ g_{78}^{(1,5)}(\omega,\omega_1,\omega_2,\mu) \right]^*
   = \int_{-\infty}^{\bar\Lambda}\!d\omega\,
    g_{78}^{(1,5)}(\omega,\omega_2,\omega_1,\mu) \,.
\end{equation}
From PT invariance of the matrix element it follows that all the shape functions are  real implying  that the functions
\begin{equation}
h_{78}^{(1,5)}  :=  \int_{-\infty}^{\bar \Lambda}  d\omega\,   g_{78}^{(1,5)}(\omega,\omega_1,\omega_2)
\end{equation}
are symmetric under the exchange of $\omega_1$ and $\omega_2$. Moreover, one also derives from the explicit form of the shape functions that  }
\begin{equation}
\int d\omega\, \bar g_{78}(\omega,\omega_1,\omega_2) = \int d\omega\, \bar g_{78}^\mathrm{cut}(\omega,\omega_1,\omega_2)\,.
\end{equation}
Thus, the contribution ${\mathcal F}_{78}^{(b)}$ vanishes. 

The other contribution is given by
\begin{align}
{\mathcal F}_{78}^{(c)} = \frac{1}{m_b}\,\frac{C_{8g}(\mu)C_{7\gamma}(\mu)}{C_{\rm OPE}}\,4\pi\alpha_s(\mu)\,
&\mathrm{Re}\int_{\frac{q_\mathrm{min}^2}{M_B}}^{\frac{q_\mathrm{max}^2}{M_B}}\frac{d\bar n\cdot q}{\bar n\cdot q}
\int d\omega_1\,d\omega_2\,\frac{1}{\omega_1-\omega_2+\bar n\cdot q+i\epsilon}\nonumber\\
\Bigg[&\Bigg(\frac{1}{\omega_1+\bar n\cdot q+i\epsilon}+\frac{1}{\omega_2-\bar n\cdot q-i\epsilon}\Bigg)h_{78}^{(1)}(\omega_1,\omega_2,\mu)\nonumber\\
-&\Bigg(\frac{1}{\omega_1+\bar n\cdot q+i\epsilon}-\frac{1}{\omega_2-\bar n\cdot q-i\epsilon}\Bigg)h_{78}^{(5)}(\omega_1,\omega_2,\mu)\Bigg]\,.
\end{align}
In the vacuum insertion approximation (see again Ref.~\cite{Benzke:2010js}) 
\begin{equation}\label{vacuumapproximation}
h_{78}^{(1)}(\omega_1,\omega_2,\mu) = h_{78}^{(5)}(\omega_1,\omega_2,\mu) = 
-e_\mathrm{spec}\,\frac{F^2(\mu)}{8}\left(1-\frac{1}{N_c^2}\right)\phi_+^B(-\omega_1,\mu)\,\phi_+^B(-\omega_2,\mu)\,,
\end{equation}
where $F=f_B\sqrt{M_B}$, $e_\mathrm{spec}$ is the charge of the $B$ meson spectator quark, and $\phi_+^B$ is the light-cone distribution amplitude (LCDA). Since the LCDAs vanish for $\omega_i\to 0$, the $\omega_i$ integrals yield
\begin{align}
-e_\mathrm{spec}\,\frac{F^2(\mu)}{8}\left(1-\frac{1}{N_c^2}\right)
(-2)\mathrm{P}\int\frac{d\omega_1}{\omega_1-\bar n\cdot q}\,\phi_+^B(-\omega_1)\,\mathrm{P}\int\frac{d\omega_2}{\omega_1-\omega_2-\bar n\cdot q}\,\phi_+^B(-\omega_2)\,.
\label{eqn:pvints}
\end{align}
In order to estimate the magnitude of this contribution we use the model for the LCDAs given in \cite{Grozin:1996pq}
\begin{equation}
\phi_+^B(\omega)=\frac{\omega}{\omega_0}e^{-\omega/\omega_0}\,,
\end{equation}
where $\omega_0=\frac{2}{3}\bar \Lambda$. Then the principal value integrals of (\ref{eqn:pvints}) can be computed analytically  and we find for the uncertainty
\begin{align}
{\mathcal F}_{78}^{(c)} = \frac{1}{m_b}\,\frac{C_{8g}(\mu)C_{7\gamma}(\mu)}{C_{\rm OPE}}\,&4\pi\alpha_s(\mu)\,e_\mathrm{spec}\,
\int_{\frac{q_\mathrm{min}^2}{M_B}}^{\frac{q_\mathrm{max}^2}{M_B}}\frac{d\bar n\cdot q}{\bar n\cdot q}
\frac{F^2(\mu)}{4}\left(1-\frac{1}{N_c^2}\right)
\frac{1}{4\omega_0^3}\nonumber\\
&\Bigg[-2\omega_0-(2\bar n\cdot q+\omega_0)e^\frac{\bar n\cdot q}{\omega_0}\mathrm{Ei}\left(-\frac{\bar n\cdot q}{\omega_0}\right)+\omega_0e^{-\frac{\bar n\cdot q}{\omega_0}}\mathrm{Ei}\left(\frac{\bar n\cdot q}{\omega_0}\right)\Bigg]\,,
\label{eqn:F78}
\end{align}
where the exponential integral is defined as 
\begin{equation}
\mathrm{Ei}(z)=-\mathrm{P}\int_{-z}^\infty\frac{e^{-t}}{t}\,dt\,.
\end{equation}
{ Using our standard set of parameters, in particular the uncertainty of the parameter $F$ (see above),  we integrate (\ref{eqn:F78}) numerically and find
\begin{equation} \label{convolution78}
{\mathcal F}_{78}^{(c)} \in \frac{1}{m_b}\frac{C_{8g}(\mu)C_{7\gamma}(\mu)}{C_{\rm OPE}}\,\, 4\pi\alpha_s(\mu)\,
e_{\rm spec}\, [0.058\,  \mathrm{GeV},0.068\,   \mathrm{GeV}]\,.\,
\end{equation}
We note that this estimate does not include any uncertainty due to the use of the VIA in Eq.~\ref{vacuumapproximation}. We can again express our numbers in  percentages:
\begin{align}
&{\mathcal F}_{78}^{(c)} \in [-0.2,-0.1]\,\%. 
\end{align}
}

\subsection{\texorpdfstring{Interference of ${\cal O}_{8g}$ with ${\cal O}_{8g}$}{Interference of Q8 with Q8}}

{The shape function $\bar g_{88}$ 
is more complicated than the ones in the previous cases, because not much is known about it.  But from the explicit form and PT invariance, one can derive that $\bar g_{88}$ is real. One can show in addition that the convolution with the hard-collinear function is real (see Ref.~\cite{Benzke:2010js}). With $\bar h_{88}:= \int  d \omega   \bar g_{88}(\omega,\omega_1,\omega_2,\mu)$ we find for the convolution integral
\begin{align} \label{O8O82}
{\cal F}_{88} &= \frac{1}{m_b} \,\frac{C_{8g}(\mu)C_{8g}(\mu)}{C_\text{OPE}}\,4\pi\alpha_s(\mu)\,e_s^2\,
\text{Re}\int_{\frac{q_\mathrm{min}^2}{M_B}}^{\frac{q_\mathrm{max}^2}{M_B}}\frac{d\bar n\cdot q}{\bar n\cdot q} \nonumber \\
    &\phantom{=\,}\,\times 
\int  \frac{d\omega_1}{\omega_1 + \bar n \cdot q + i \epsilon}\,  \frac{d \omega_2}{\omega_2 +\bar n \cdot q - i \epsilon}\, 2\bar h_{88} (\omega_1,\omega_2,\mu) \,.
\end{align}
We cannot get any stricter estimation from the convolution, however we have been able to separate factors like $e_s^2$ etc, thus we estimate
\begin{align}
\Lambda(\mu) =&\,\text{Re}\int_{\frac{q_\mathrm{min}^2}{M_B}}^{\frac{q_\mathrm{max}^2}{M_B}}\frac{d\bar n\cdot q}{\bar n\cdot q}
\int  \frac{d\omega_1}{\omega_1 + \bar n \cdot q + i \epsilon}\,  \frac{d \omega_2}{\omega_2 +\bar n \cdot q - i \epsilon}\, 2\bar h_{88} (\omega_1,\omega_2,\mu)
\end{align}
to be of ${\cal O}(\Lambda_\text{QCD})$. So we assume $0\,\text{GeV}<\Lambda(\mu)<1\,\text{GeV}$\footnote{As mentioned below Eq.~\ref{eq:g88def} there is a subtlety concerning the convolution integral in Eq.~\ref{O8O82}. We state here that the logarithmic dependence on the parameter $\Lambda_{{\rm UV}}$ is assumed to be included in our hadronic parameter $\Lambda(\mu)$, which is therefore independent of $\Lambda_{{\rm UV}}$.}.

Compared to the estimates found in Eqs.~(\ref{convolution17}) and (\ref{convolution78}), this leads to a rather conservative estimate of the convolution integral
\begin{align}
&{\mathcal F}_{88} \in [0,0.5]\,\%\,.
\end{align}

\subsection{Summary of the numerical analysis} 

Our estimates of the resolved contributions to the leading order
 in $1/m_b$, 
\begin{equation}
{\mathcal F}_{17}\in [-0.5,+3.4]\,\%,\,\,\,\,\,
{\mathcal F}_{78}  \in [-0.2,-0.1]\,\%,\,\,\,\,\,
{\mathcal F}_{88}  \in [0,0.5]\,\%\,.
\end{equation}
can be now summed up using the scanning method. Our final result is
\begin {align} 
 {\mathcal F}_{{1/m_b}}\in [-0.7,+3.8]\,.
\end{align}
As discussed, this estimate of the resolved contributions  represents an irreducible theoretical  uncertainty of the total decay  rate of the inclusive $\bar B \to X_s \ell^+\ell^-$. The results in 
Section~\ref{sec:contributions}  allow to make analogous estimates for the other two independent 
angular observables within  the inclusive $\bar B \to X_s \ell^+\ell^-$.


\section{Conclusions} \label{sec:conclusion}
{The present and future measurements of the inclusive decay 
$\bar B \rightarrow X_s \ell^+\ell^-$ need a hadronic mass cut in order to suppress potential huge background. The cut on the hadronic mass implies specific kinematics in which the standard local OPE breaks down and non-perturbative $b$-quark distributions, so-called shape functions, have to be introduced. The specific kinematics of  low dilepton masses $q^2$  and small hadronic mass $M_X$ leads to a multi-scale problem for which soft-collinear effective theory is  the appropriate tool.

In this paper, we  have identified  the correct power counting of all variables in the low-$q^2$ window of the inclusive decay $\bar B \rightarrow X_s \ell^+\ell^-$  within the effective theory SCET  if such an hadronic mass  cut is imposed. We  have analysed the resolved power corrections at the  order $1/m_b$ in a systematic way. Resolved  contributions are those in which the virtual photon couples to light partons instead of connecting directly to the effective weak-interaction vertex. They stay non-local even if  the hadronic mass cut is released. Thus, they represent an irreducible uncertainty independent of the hadronic mass cut.

We have presented numerical estimates of the corresponding uncertainties to the first order in $1/m_b$. 
We find an overall uncertainty of  ${\mathcal F}_{{1/m_b}}\in [-0.7,+3.8]$ for the decay rate. 
Numerical estimates of the uncertainties in the case of the other two independent angular observables in the inclusive decay $\bar B \rightarrow X_s \ell^+\ell^-$  can be easily derived from the analytical results of this paper.
However, we have found indications that the subleading contributions to order $1/m_b^2$ might be numerically relevant due to the large ratio  $C_9/C_{7\gamma}$ which calls for an additional calculation~\cite{workinprogress}.}


\acknowledgments

We thank Tobias Huber for valuable help and discussions. TH thanks the CERN theory group for its hospitality during his regular visits to CERN where part of this work was written. We thank Michael Fickinger for crucial input  at an early stage of the project.


\newpage

\begin{thebibliography}{99}

\bibitem{Isidori:2010kg}
  G.~Isidori, Y.~Nir and G.~Perez,
  Ann.\ Rev.\ Nucl.\ Part.\ Sci.\  {\bf 60} (2010) 355
  [arXiv:1002.0900 [hep-ph]].


\bibitem{Hurth:2003vb}
  T.~Hurth,
  Rev.\ Mod.\ Phys.\  {\bf 75} (2003) 1159
  [arXiv:hep-ph/0212304].

\bibitem{Hurth:2010tk}
  T.~Hurth and M.~Nakao,
  Ann.\ Rev.\ Nucl.\ Part.\ Sci.\  {\bf 60} (2010) 645
  [arXiv:1005.1224 [hep-ph]].

\bibitem{Hurth:2012vp}
  T.~Hurth and F.~Mahmoudi,
  Rev.\ Mod.\ Phys.\  {\bf 85} (2013) 795
  [arXiv:1211.6453 [hep-ph]].



\bibitem{Huber:2015sra} 
  T.~Huber, T.~Hurth and E.~Lunghi,
  JHEP {\bf 1506}, 176 (2015)
  doi:10.1007/JHEP06(2015)176
  [arXiv:1503.04849 [hep-ph]].






\bibitem{Aaij:2013qta} 
  R.~Aaij {\it et al.} [LHCb Collaboration],
  Phys.\ Rev.\ Lett.\  {\bf 111}, 191801 (2013)
  doi:10.1103/PhysRevLett.111.191801
  [arXiv:1308.1707 [hep-ex]].

\bibitem{Aaij:2015oid} 
  R.~Aaij {\it et al.} [LHCb Collaboration],
  JHEP {\bf 1602}, 104 (2016)
  doi:10.1007/JHEP02(2016)104
  [arXiv:1512.04442 [hep-ex]].

\bibitem{Hurth:2013ssa} 
  T.~Hurth and F.~Mahmoudi,
  JHEP {\bf 1404}, 097 (2014)
  doi:10.1007/JHEP04(2014)097
  [arXiv:1312.5267 [hep-ph]].



\bibitem{Hurth:2014zja}
  T.~Hurth and F.~Mahmoudi,
  arXiv:1411.2786 [hep-ph].



\bibitem{Iwasaki:2005sy}
  M.~Iwasaki {\it et al.}  [Belle Collaboration],
  hep-ex/0503044.

\bibitem{Lees:2013nxa}
  J.~P.~Lees {\it et al.}  [BaBar Collaboration],
  Phys.\ Rev.\ Lett.\  {\bf 112} (2014) 211802
  [arXiv:1312.5364 [hep-ex]].

\bibitem{Aubert:2004it}
B.~Aubert {\it et al.}  [BABAR Collaboration],
Phys.\ Rev.\ Lett.\  {\bf 93} (2004) 081802
[hep-ex/0404006].

\bibitem{Sato:2014pjr}
  Y.~Sato {\it et al.}  [Belle Collaboration],
  arXiv:1402.7134 [hep-ex].

\bibitem{Belle2}
  T.~Abe {\it et al.}  [Belle-II Collaboration],
  arXiv:1011.0352 [physics.ins-det].








\bibitem{Benzke:2010js}
  M.~Benzke, S.~J.~Lee, M.~Neubert and G.~Paz,
  JHEP {\bf 1008} (2010) 099
  [arXiv:1003.5012 [hep-ph]].

\bibitem{Benzke:2010tq}
  M.~Benzke, S.~J.~Lee, M.~Neubert and G.~Paz,
  Phys.\ Rev.\ Lett.\  {\bf 106} (2011) 141801
  [arXiv:1012.3167 [hep-ph]].




\bibitem{Chay:1990da}
  Chay J, Georgi H, Grinstein B. 
     Phys.\ Lett.\  B {\bf 247} (1990) 399.
 
 \bibitem{Bigi:1992su}
  Bigi II, Uraltsev NG, Vainshtein AI.
     Phys.\ Lett.\  B {\bf 293} (1992) 430
   [Erratum-ibid.\  B {\bf 297} (1993) 477]
   [arXiv:hep-ph/9207214].
  





\bibitem{Isgur:1991xa}
       N.~Isgur and M.~B.~Wise,
  Adv.\ Ser.\ Direct.\ High Energy Phys.\  {\bf 10}, 549 (1992).
  
\bibitem{Neubert:1993mb}
     M.~Neubert,
  Phys.\ Rept.\  {\bf 245} (1994) 259
   [arXiv:hep-ph/9306320].



\bibitem{Falk:1993dh}
       A.~F.~Falk, M.~E.~Luke and M.~J.~Savage,
     Phys.\ Rev.\  D {\bf 49} (1994) 3367
   [arXiv:hep-ph/9308288].

\bibitem{Ali:1996bm}
    A.~Ali, G.~Hiller, L.~T.~Handoko and T.~Morozumi,
     Phys.\ Rev.\  D {\bf 55} (1997) 4105
   [arXiv:hep-ph/9609449].

\bibitem{Chen:1997dj}
       J.~W.~Chen, G.~Rupak and M.~J.~Savage,
     Phys.\ Lett.\  B {\bf 410} (1997) 285
   [arXiv:hep-ph/9705219].

\bibitem{Buchalla:1998mt}
       G.~Buchalla and G.~Isidori,
     Nucl.\ Phys.\  B {\bf 525}, 333 (1998)
   [arXiv:hep-ph/9801456].

\bibitem{Bauer:1999kf}
       C.~W.~Bauer and C.~N.~Burrell,
     Phys.\ Rev.\  D {\bf 62}, 114028 (2000)
   [arXiv:hep-ph/9911404].

\bibitem{Mannel:2010wj}
  T.~Mannel, S.~Turczyk and N.~Uraltsev,
  JHEP {\bf 1011} (2010) 109
  doi:10.1007/JHEP11(2010)109
  [arXiv:1009.4622 [hep-ph]].


\bibitem{Ligeti:1997tc}
    Z.~Ligeti, L.~Randall and M.~B.~Wise,
     Phys.\ Lett.\ B {\bf 402}, 178 (1997)
   [hep-ph/9702322].

\bibitem{Grant:1997ec} 
  A.~K.~Grant, A.~G.~Morgan, S.~Nussinov and R.~D.~Peccei,
  Phys.\ Rev.\ D {\bf 56}, 3151 (1997)
  doi:10.1103/PhysRevD.56.3151
  [hep-ph/9702380].





\bibitem{Buchalla:1997ky} 
  G.~Buchalla, G.~Isidori and S.~J.~Rey,
  Nucl.\ Phys.\ B {\bf 511}, 594 (1998)
  [hep-ph/9705253].






\bibitem{Voloshin:1996gw}
   M.~B.~Voloshin,
     Phys.\ Lett.\ B {\bf 397}, 275 (1997)
   [hep-ph/9612483].

\bibitem{Lee:2006wn}
     S.~J.~Lee, M.~Neubert and G.~Paz,
     Phys.\ Rev.\  D {\bf 75} (2007) 114005
   [arXiv:hep-ph/0609224].


\bibitem{Lee:2005pk}
     K.~S.~M.~Lee and I.~W.~Stewart,
     Phys.\ Rev.\ D {\bf 74} (2006) 014005
   [hep-ph/0511334].

\bibitem{Lee:2005pw}
       K.~S.~M.~Lee, Z.~Ligeti, I.~W.~Stewart and F.~J.~Tackmann,
     Phys.\ Rev.\ D {\bf 74} (2006) 011501
   [hep-ph/0512191].

\bibitem{Lee:2008xc}
      K.~S.~M.~Lee and F.~J.~Tackmann,
     Phys.\ Rev.\  D {\bf 79} (2009) 114021
   [arXiv:0812.0001 [hep-ph]].




\bibitem{Kruger:1996cv}
      F.~Kruger and L.~M.~Sehgal,
     Phys.\ Lett.\  B {\bf 380}, 199 (1996)
   [arXiv:hep-ph/9603237].

\bibitem{Kruger:1996dt}
     F.~Kruger and L.~M.~Sehgal,
     Phys.\ Rev.\  D {\bf 55} (1997) 2799
   [arXiv:hep-ph/9608361].






\bibitem{Buchalla:1995vs}
  G.~Buchalla, A.~J.~Buras and M.~E.~Lautenbacher,
  Rev.\ Mod.\ Phys.\  {\bf 68} (1996) 1125
  [hep-ph/9512380].

\bibitem{Beneke:2001at}
  M.~Beneke, T.~Feldmann and D.~Seidel,
  Nucl.\ Phys.\ B {\bf 612} (2001) 25
  doi:10.1016/S0550-3213(01)00366-2
  [hep-ph/0106067].
  
  





\bibitem{Beneke:2009az}
    M.~Beneke, G.~Buchalla, M.~Neubert and C.~T.~Sachrajda,
   Eur.\ Phys.\ J.\  C {\bf 61} (2009) 439
   [arXiv:0902.4446 [hep-ph]].





\bibitem{Bauer:2001yt}
  C.~W.~Bauer, D.~Pirjol and I.~W.~Stewart,
  Phys.\ Rev.\  D {\bf 65}, 054022 (2002)
  [arXiv:hep-ph/0109045].

\bibitem{Beneke:2002ph}
  M.~Beneke, A.~P.~Chapovsky, M.~Diehl and T.~Feldmann,
  Nucl.\ Phys.\ B {\bf 643} (2002) 431
  [hep-ph/0206152].

\bibitem{Beneke:2002ni}
  M.~Beneke and T.~Feldmann,
  Phys.\ Lett.\ B {\bf 553} (2003) 267
  [hep-ph/0211358].






\bibitem{Bauer:2000yr}
  C.~W.~Bauer, S.~Fleming, D.~Pirjol and I.~W.~Stewart,
  Phys.\ Rev.\ D {\bf 63} (2001) 114020
  doi:10.1103/PhysRevD.63.114020
  [hep-ph/0011336].

\bibitem{Bauer:2003mga}
  C.~W.~Bauer, D.~Pirjol and I.~W.~Stewart,
  Phys.\ Rev.\ D {\bf 68} (2003) 034021
  doi:10.1103/PhysRevD.68.034021
  [hep-ph/0303156].

\bibitem{Chetyrkin:1996vx}
  K.~G.~Chetyrkin, M.~Misiak and M.~Munz,
  Phys.\ Lett.\ B {\bf 400} (1997) 206
   Erratum: [Phys.\ Lett.\ B {\bf 425} (1998) 414]
  doi:10.1016/S0370-2693(97)00324-9
  [hep-ph/9612313].



\bibitem{Bosch:2004th}
  S.~W.~Bosch, B.~O.~Lange, M.~Neubert and G.~Paz,
  Nucl.\ Phys.\  B {\bf 699}, 335 (2004)
  [arXiv:hep-ph/0402094].





\bibitem{Grozin:1996pq}
  A.~G.~Grozin and M.~Neubert,
  Phys.\ Rev.\ D {\bf 55} (1997) 272
  doi:10.1103/PhysRevD.55.272
  [hep-ph/9607366].
  
\bibitem{workinprogress}
M.~Benzke, T.~Hurth and S.~Turczyk, work in progress.  


\end{thebibliography}
\end{document}